\documentclass[conference]{IEEEtran}
\IEEEoverridecommandlockouts
\usepackage{multirow}
\usepackage{tablefootnote}
\usepackage{enumitem}
\usepackage{amsmath}
\usepackage{xcolor}

\usepackage{soul}
\usepackage{booktabs}
\usepackage{colortbl}
\usepackage{color}
\usepackage{array}

\usepackage{ifpdf}
\usepackage{amsmath, nccmath,amssymb,amsfonts,amsthm}
\usepackage{lipsum}

\usepackage{cite}

\usepackage{algorithmic}
\usepackage{textcomp}
\usepackage{graphicx}
\usepackage{amsfonts}
\usepackage{todonotes}

\begin{document}

\title{MISS: Multi-Interest Self-Supervised Learning Framework for Click-Through Rate Prediction}

\author{\IEEEauthorblockN{Wei Guo$^{1}$, Can Zhang$^{2}$, Zhicheng He$^{1}$, Jiarui Qin$^{3}$, Huifeng Guo$^{1}$, Bo Chen$^{1}$, \\ Ruiming Tang$^{1}$, Xiuqiang He$^{1}$, Rui Zhang$^{4}$}
\IEEEauthorblockA{$^{1}$Huawei Noah's Ark Lab, $^{2}$National University of Singapore\\
$^{3}$Shanghai Jiao Tong University,
$^{4}$www.ruizhang.info\\
\{guowei67, hezhicheng9, huifeng.guo, chenbo116, tangruiming, hexiuqiang1\}@huawei.com\\
can.zhang@u.nus.edu, qinjr@apex.sjtu.edu.cn,
rayteam@yeah.net
}
}

\maketitle
\begin{abstract}
CTR prediction is essential for modern recommender systems.
Ranging from early factorization machines to deep learning based models in recent years, existing CTR methods focus on capturing useful feature interactions or mining important behavior patterns.
Despite the effectiveness, we argue that these methods suffer from the risk of \textit{label sparsity} (i.e., the user-item interactions are highly sparse with respect to the feature space), \textit{label noise} (i.e., the collected user-item interactions are usually noisy), and the underuse of domain knowledge (i.e., the pairwise correlations between samples). 
To address these challenging problems, we propose a novel Multi-Interest Self-Supervised learning (MISS) framework which enhances the feature embeddings with interest-level self-supervision signals.
With the help of two novel CNN-based multi-interest extractors, self-supervision signals are discovered with full considerations of different \textit{interest representations} (point-wise and union-wise), \textit{interest dependencies} (short-range and long-range), and \textit{interest correlations} (inter-item and intra-item).
Based on that, contrastive learning losses are further applied to the augmented views of interest representations, which effectively improves the feature representation learning.
Furthermore, our proposed MISS framework can be used as an ``plug-in'' component with existing CTR prediction models and further boost their performances.
Extensive experiments on three large-scale datasets show that MISS significantly outperforms the state-of-the-art models, by up to 13.55\% in \textit{AUC}, and also enjoys good compatibility with representative deep CTR models.
\noindent\let\thefootnote\relax\footnotetext{*Wei Guo, Can Zhang and Zhicheng He are co-first authors with equal contributions. Ruiming Tang and Rui Zhang are the co-corresponding authors.}
\end{abstract}

\IEEEpeerreviewmaketitle

\vspace{-3mm}
\section{Introduction}\label{sec:introduction}
Click-Through Rate (CTR) prediction is an essential task in the domain of online advertising and recommender systems, both of which are multi-billion dollar businesses nowadays.
As shown in Table \ref{tab:example}, the data involved in CTR prediction are mostly in a multi-field tabular format. Each row represents a sample\footnote{In conventional practices, a sample is made up by a given user, a candidate item, and other associated information including profiles and behavior histories.}
described by multiple \textit{fields}\footnote{The instantiation of a field is a \textit{feature}. 
When there is no ambiguity, we use ``field" and ``feature" exchangeably.} such as gender, click history, item ID, and item category.
CTR prediction is to estimate the probability that a user will click an item based on the multi-field features.
Due to the powerful feature representation learning ability, the mainstream of CTR prediction research is dominated by deep learning models~\cite{cheng2016wide, guo2017deepfm, zhou2018deep}.
Deep CTR prediction models have made great progresses and have been deployed in many commercial recommender systems, such as Wide\&Deep~\cite{cheng2016wide} in Google Play, DeepFM~\cite{guo2017deepfm} in Huawei AppGallery, and Deep Interest Network (DIN)~\cite{zhou2018deep} in Taobao. 

Despite the great successes, CTR prediction models are all faced with the \textit{label sparsity} and \textit{label noise} problems, which deteriorate quickly with the rapid growth of data volume and feature size in online systems.
What is more, existing approaches also \textit{underuse the domain knowledge} implicitly contained in the data. 
Without solving the above three problems, it is difficult for existing CTR prediction models to learn effective feature representations with the sparse and noisy user-item interactions which serve as the supervision signals.
In this work, we seek to utilize self-supervised learning (SSL) to solve the above mentioned three problems.
On the one hand, the self-supervision signals are extracted based on the understanding of recommendation domain knowledge, which can effectively supplement the original sparse supervision signals. On the other hand, the self-supervision loss also regularizes the learned representations and filters out noises.
\begin{table*}[t]
\vspace{-5mm}
\setlength{\abovecaptionskip}{-0.1cm}
\setlength{\belowcaptionskip}{-0.1cm}
\caption{An example of multi-field data for CTR prediction, where each row indicates a sample, and each column represents a field.}
\centering
\begin{tabular}{ccccccc|c} \hline\hline
User  & Gender  & City          & Click History                     & Item           & Item Category&Day   & Click  \\ \hline
Lisa   & Female  & New York      & Honor50, iPhone12, MI11           & Mate40 Pro     & Cellphone    &Mon.  & 1 \\
David  & Male    & Los Angeles   & Diaper, Milk powder, Shave cream  & Draft beer     & Beverage     &Sat.  & 1  \\
Yakov  & Male    & Moscow        & Caviar, Vodka, Dark chocolate     & Whisky         & Wine         &Sun.  & 0\\
Meimei & Female  & Beijing       & Umbrella, Chopsticks, Detergent   & Running shoes  & Outdoor      &Fri.  & 0 \\
Eliza  & Female  & London        & Sun glasses, Boots, Wind-breaker  & Heels          & Shoes        &Fri.  & 1 \\
Yoshida & Male   & Tokyo         & Salmon, Sea urchin, Wasabi        & Beer           & Beverage     &Wed.  & 1 \\ 
\hline\hline
\end{tabular}
\label{tab:example}
\vspace{-5mm}
\end{table*}

\begin{figure}[t]
\centering
\setlength{\abovecaptionskip}{-0.2cm}
\noindent\makebox[0.45\textwidth][c]{\includegraphics[scale=0.25]{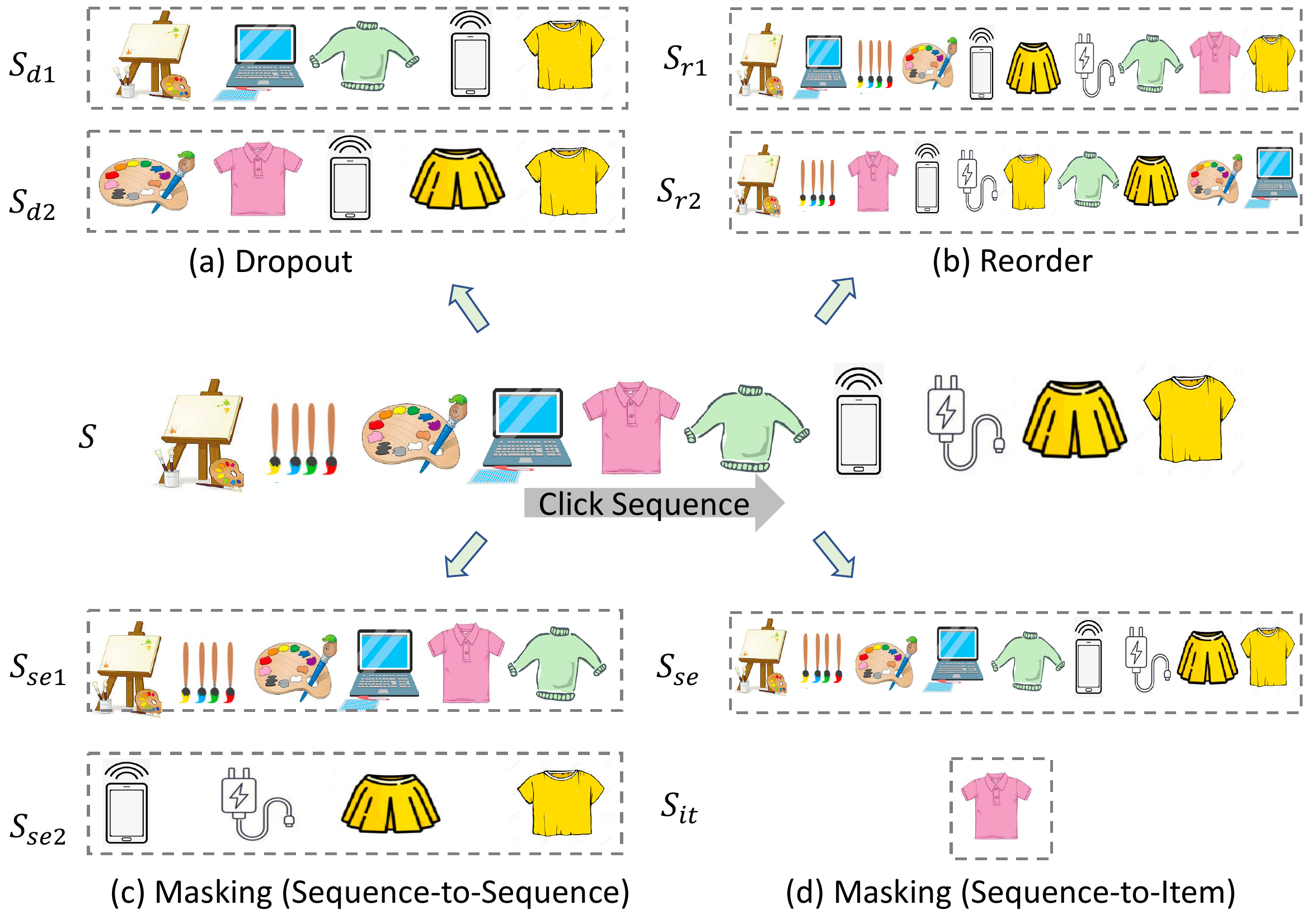}}
\caption{Data augmentation in existing SSL-based recommendation models. (a) Dropout, (b) Reorder, (c) Sequence-to-Sequence Masking, (d) Sequence-to-Item Masking.}
\label{fig:teaser}
\vspace{-5mm}
\end{figure}

The basic framework for SSL mainly contains two key components: data augmentation for enhancing training data and contrastive losses for enhancing supervision signals.
Though making great progress in both CV~\cite{chen2020simple, gidaris2018unsupervised, henaff2021efficient} and NLP~\cite{devlin2018bert,Lan2020ALBERT, liu2019roberta} fields, SSL has not been fully explored in recommendation tasks.
As illustrated in Figure \ref{fig:teaser}, current SSL-based CTR models generally adopt three kinds of augmentation operators, i.e., \textit{dropout}, \textit{reorder}, and \textit{masking}, all of which are directly introduced from CV or NLP areas without appropriate adaptation to recommendation tasks.
After augmentation, each behavior sequence is transformed into two different new sequences (i.e., a pair of views) which are required to be similar in the contrastive learning stage.
However, in the recommendation domain, it is natural that user behaviors are of \textit{multi-interest}, as stated in \cite{zhou2018deep,Li2019Multi,Cen2020Controllable}, so one augmented pair of views may contain very different interests even when they are obtained from the same user behavior sequence.
As a consequence, maximizing the pairwise similarities in contrastive learning may introduce noises and deteriorate representation learning in recommendation tasks.

\begin{figure}[t]
\centering
\setlength{\abovecaptionskip}{-0.2cm}
\noindent\makebox[0.45\textwidth][c]{\includegraphics[scale=0.3]{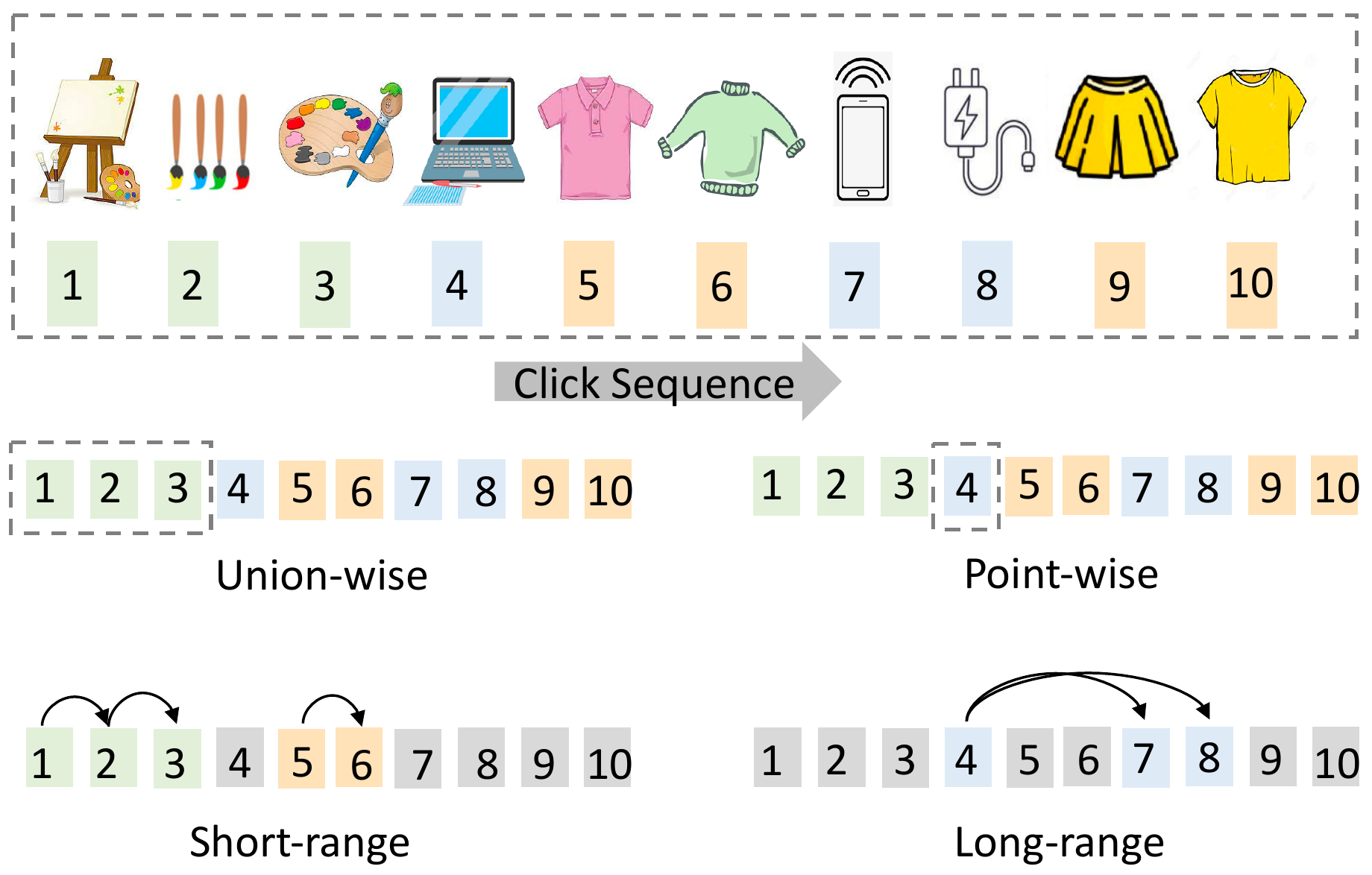}}
\caption{A click behavior sequence with multiple user interests.} 
\label{fig:interest}
\vspace{-5mm}
\end{figure}

In this paper, we study the self-supervised learning for CTR prediction with the consideration of multi-interest.
To incorporate the multi-interest in user behavior sequences, we design three important interest modeling practices to better utilize the domain knowledge.
\begin{itemize}[leftmargin=*]
\item \textbf{Point-wise and Union-wise Interest Representations.}
Within a user behavior sequence, an interest can not only be represented as a single behavior in a point-wise manner, but also can be represented as several behaviors in a union-wise manner.
For example, as shown in Figure~\ref{fig:interest}, the paint board, the brush, and the palette together represent user's interest in painting tools, while notebook alone is enough to indicate an interest in electronic products.
Therefore, it is important to simultaneously consider both point-wise and union-wise interest representations to provide more sufficient self-supervision signals.


\item \textbf{Short-range and Long-range Interest Dependencies.}
It is possible that user behaviors of one interest are interleaved with behaviors of another interest, in which case modeling long-range dependencies is necessary. For instance, in Figure~\ref{fig:interest}, behaviors on electronic products (i.e., computer, phone, and charger) are interleaved by behaviors on clothes (i.e., red T-shirt and green sweater). 
It is also common that behaviors of an interest are consecutive without any interruption, e.g., the first three behaviors of painting tools in Figure~\ref{fig:interest}, thus short-range dependencies also need to be learned.
Therefore, mining long-range and short-range dependencies are complementary when modeling user behavior sequences with multiple interests. 
\item \textbf{Inter-item and Intra-item Interest Correlations.}
Besides the inter-item interest correlations discussed above, the interest correlations between different item attributes (defined as intra-item correlation) also contain useful self-supervision signals. For examples, some people like Nike sneakers while some other people may prefer cheap slippers. 
Therefore, it is necessary to extract self-supervision signals from both inter-item and intra-item correlations together.
\end{itemize}

To this end, we propose a novel Multi-Interest Self-Supervised learning (MISS) framework for deep CTR models.
To mine self-supervision signals from user behaviors of multi-interests, MISS proposes interest-level contrastive losses to take the place of sample-level losses.
Specifically, individual self-supervision signals are extracted for multiple interests from the user behaviors.
By means of Convolutional Neural Network (CNN), both point-wise and union-wise interest representations are learned from the local correlations of behaviors on the time line, and the short-range and long-range interest dependencies are extracted by considering different distances of interest representations.
Finally, the intra-item correlations are also modeled by sampling different feature combinations with convolution kernels.
Furthermore, MISS serves as a model-agnostic embedding learning framework for user behavior sequence features, which is able to work compatibly with the existing deep CTR models, including methods based on both feature interactions and user interest mining.
 
To summarize, our work makes three major contributions as follows:
\begin{enumerate}[leftmargin=*]
    \item We propose a novel Multi-Interest Self-Supervised Learning framework named MISS, which enhances feature embeddings in an end-to-end manner.
    As far as we know, our work is the first to apply interest-level contrastive losses for recommendation tasks. 
    \item More specially, we propose CNN-based self-supervision signal extractors with full considerations of different interest representations (point-wise and union-wise), interest dependencies (short-range and long-range) and interest correlations (inter-item and intra-item).
    Based on that, contrative learning is implemented to  make better use of the domain knowledge and to make the best of interest-level self-supervision knowledge.
    \item Extensive experiments demonstrate that our MISS framework not only achieves state-of-the-art performances on three large-scale datasets, but also enjoys excellent compatibility with various representative baselines. 
\end{enumerate}

\section{Related work}\label{sec:related_works}
\subsection{Deep CTR Models}
According to different model architectures, recent CTR models can be divided into two categories: feature interaction based models and user interest modeling based models. In the following, we give a brief introduction about these two kinds of models \cite{zhang2016deep, cheng2016wide, guo2017deepfm,wang2017deep,guo2019order,chen2021enhancing,Guo2021Dual,su2021detecting,su2021neural}, interested readers can refer to the recent survey paper \cite{Zhang2021Deep} for more details.

Feature interaction based models focus on learning sophisticated interactions between different features, and the representative models include 
Wide\&Deep~\cite{cheng2016wide}, DeepFM~\cite{guo2017deepfm}, DCN~\cite{wang2017deep} and DCN-M\cite{wang2021dcn}. 
Wide\&Deep~\cite{cheng2016wide} learning builds a wide linear component and a DNN component to model explicit and implicit feature interactions respectively.
However, manual efforts for feature engineering are still required in its wide component. 
To avoid such manual efforts, DeepFM~\cite{guo2017deepfm} is thus proposed to replace the wide part with FM and share the input features between deep and wide components. 
DCN~\cite{wang2017deep} explicitly and automatically applies feature crossing for improving accuracy and efficiency of the DNN model.
DCN-M\cite{wang2021dcn} further improves DCN by replacing the cross vector into a cross matrix to enhance it's learning ability.

On the other hand, user interest modeling based models dedicate to capture important patterns from sequential behavior fields.
The mainstream models include DIN~\cite{zhou2018deep}, DIEN~\cite{zhou2019deep}, and DSIN~\cite{Feng2019Deep} which aim to use auto-regressive models to learn users' diverse interests precisely.
DIN~\cite{zhou2018deep} proposes a local activation unit to adaptively learn candidate-wise user interest representations from the diverse behavior sequences, based on which the CTR score is estimated.
Based on DIN, DIEN~\cite{zhou2019deep} further proposes to capture the interest evolving process with an auxiliary loss, thus better deals with interest drifting.
Considering the homogeneity of user behaviors within each session, DSIN~\cite{Feng2019Deep} adopts self-attention and Bi-LSTM to capture intra- and inter-session interest representations respectively. 
To retrieve more relevant user behavior interests from long history sequences, search-based models like SIM~\cite{Pi2020Search} and UBR4CTR~\cite{qin2020user} have also been proposed.
To better utilize the user-item relevance, DMR~\cite{lyu2020deep} adopts the attention networks to learn user and item representations from the user-item and item-item interaction networks.

Despite the great progresses achieved by feature interaction and user interest modeling models, there are  three common problems hindering the performances of both lines of approaches, i.e., label sparsity, label noise and the underuse of domain knowledge. 
To tackle these three problems, we propose a self-supervised learning framework tailored for deep CTR models. Through data augmentation and interest-level contrastive learning, self-supervision signals and pairwise  correlations  between  samples can be utilized to enhance user interest representations learned from sparse and noisy data.
What is more, our proposed framework is model-agnostic which can be seamlessly applied to both feature interaction and user behavior modeling approaches, as described and verified in the following sections.

\subsection{Self-Supervised Learning}
Self-supervised learning~\cite{he2020momentum, chen2020simple, caron2018deep, chen2020big} has recently become an emerging trend in CV and NLP areas.
SSL models enhance the learned representations with self-supervision signals extracted from unlabeled data, thus alleviating deep models' heavy dependence on manual labels.
According to the model architectures and learning objectives~\cite{liu2020self}, SSL models can be categorized into two genres, i.e., generative models and contrastive models.

Generative SSL models exploit context features by modeling the generation processes, and typical examples include the BERT models~\cite{devlin2018bert, Joshi2020SpanBERT, Lan2020ALBERT}. 
On the other hand, contrastive SSL models utilize discrimination information in a ``learn to compare" manner, which maximizes the correlation between similar instances~\cite{gutmann2010noise, oord2018representation}. 
Following this paradigm, Deep InfoMax (DIM)~\cite{hjelm2018learning} explicitly learns the Mutual Information Maximization (MIM) objective between features from the local patches and the whole input image.
Similarly, Contrastive Predictive Coding (CPC)~\cite{oord2018representation} learns MIM between audio segments and their context audios. 
Deep Graph InfoMax (DGI)~\cite{velivckovic2018deep} explicitly maximizes the correlation between a node and its 2-hop neighbors in the context graph with MIM.


Despite the successes achieved in CV and NLP areas, exploiting SSL in recommendation is still an under-explored task where few works have been proposed so far. 
To characterize the intrinsic data correlations, S3Rec~\cite{zhou2020s3} combines SSL with sequential recommendation by utilizing four MIM objectives, i.e., Item-Attribute MIM, Sequence-Item MIM, Sequence-Attribute MIM, and Sequence-Sequence MIM.
In large-scale item recommendations, an auxiliary SSL task is employed to explore feature correlations by applying different feature masking patterns~\cite{yao2021selfsupervised}. 
CL4SRec~\cite{xie2020contrastive} proposes three data augmentation techniques (i.e., cropping, masking and reordering) from which two methods are randomly sampled and applied to each user sequence. 
Instead of performing self-supervision in the data space, \cite{ma2020disentangled} proposes a sequence-to-sequence training strategy to extract extra supervision signals from pairwise sub-sequences in the disentangled latent space. SGL~\cite{wu2020self} extends SSL to GCN-based recommendation models by augmenting ID embeddings and graph structures. 

However, the above SSL recommendation models directly borrow the ideas from CV and NLP areas without carefully considering the characteristics of recommendation tasks, especially the interest diversity of individual users. 
In consequence, two instances generated from the same user behavior sequence are unconditionally required to be similar in contrastive learning, even if they are derived from different interests. 
Such one-size-fits-all practices inevitably introduce noises into representation learning, thus harm the recommendation performance. 
To this end, we propose a new SSL CTR framework to incorporate self-supervision signals at the interest level.
By considering historical behavior dependencies under multiple interests, user representations are enhanced by better exploiting the intra-interest behavioral self-supervision while avoiding inter-interest contrastive learning.

\section{PRELIMINARY}\label{sec:preliminary}
For the ease of understanding, in this section, we begin by the formal definition of the CTR prediction task and necessary notations, followed by the limitation analysis of current deep CTR models and the outline of this work.

\subsection{Problem Formulation}
In a recommender system, data samples are usually stored in a multi-field format as shown in Table \ref{tab:example}. 
For CTR prediction purpose, necessary features are retrieved and combined in a fixed format to describe each sample. For example, sample of user Yoshida can be represented as
$$
\scriptsize{
\begin{matrix}
 \underbrace{[\text{Yoshida}]} & \underbrace{[\text{Male,\ ...}]} & \underbrace{[\text{Salmon,\ ...}]} & ... & \underbrace{[\text{Beer}]} & \underbrace{[\text{Beverage}]} & \underbrace{[\text{Fri.}]} \\
 \texttt{User} & \texttt{Profile} & \texttt{Click}\ \texttt{Seq.} & ... & \texttt{Item} & \texttt{Item}\ \texttt{Cate.} & \texttt{Context}
\end{matrix}
}
$$
Based on the collected data, CTR prediction is to estimate the probability that a user (i.e., Yoshida) will click a candidate item (i.e., Beer) under the given context (i.e., Friday).

In symbolic language, suppose there are $|\mathcal{U}|$ users $\mathcal{U} = \left\{u_1,u_2,...,u_{|\mathcal{U}|}\right\}$ and $|\mathcal{V}|$ items $\mathcal{V} = \left\{v_1,v_2,...,v_{|\mathcal{V}|}\right\}$, each user or item is accompanied with some attribute information such as user gender and item category, and the behavior sequence of each user is also collected and chronologically ordered as $\textbf{b} = \{v_1,v_2,\cdots,v_L\}$, where $v_l$ is the $l$-th interacted item and $L$ is the sequence length.
Thus a long raw feature vector is constructed for each sample $x$ through the combination of categorical and sequential features:
\begin{equation}\label{equ:fea_vec}
\setlength{\abovedisplayskip}{3pt}
\setlength{\belowdisplayskip}{3pt}
\textbf{x} = [f_1, \cdots, f_i, \cdots, f_I,  \textbf{s}_1, \cdots, \textbf{s}_j,  \cdots, \textbf{s}_J],
\end{equation}
where $f_i$ is a categorical feature, $\textbf{s}_j$ is a sequential feature, $I$ and $J$ denote the numbers of categorical and sequential features respectively.
Note that, besides the item ID sequence, the attribute sequences of interacted items are also useful for CTR prediction such as the category sequence $\textbf{c} = \{c_1,c_2,\cdots,c_L\}$ and price sequence $\textbf{p} = \{p_1,p_2,\cdots,p_L\}$, thus we have $\textbf{s}_j \in \{\textbf{b}, \textbf{c}, \textbf{p}\}$.
However, the sequential features are not limited to these three kinds, but can also incorporate other features according to specific tasks.
Take $\textbf{x}$ as input, a CTR model is learned to minimize the following loss function:
\begin{equation}\label{equ:obj}
\setlength{\abovedisplayskip}{3pt}
\setlength{\belowdisplayskip}{3pt}
\mathop{\min}\limits_{\Theta}\sum_{(x, y)}{\Delta(y, \text{CTRModel}(\textbf{x}; \Theta))},
\end{equation}
where $y\in \{0,1\}$ is the ground-truth click label, $\text{CTRModel}(\cdot,\cdot)$ is the CTR model with parameter set $\Theta$, and $\Delta(\cdot, \cdot)$ is the loss function. 

\subsection{Limitation Analysis}
To tackle the CTR prediction task described in the last subsection, various machine learning approaches have been proposed as described in Section \ref{sec:related_works}.
Despite the achievements, there are two common problems that seriously influence the performances of existing CTR models, i.e., \textit{label sparsity} and \textit{label noise}. Here we explain these two critical problems in details.

\begin{itemize}[leftmargin=*]
\item \textbf{Label Sparsity.} 
The observed user-item interactions, which serve as supervision signals for CTR models, are highly \textit{sparse} with respect to the number of items~\cite{qu2018product, shalev2017failures}.
Moreover, there are numerous cold-start users and infrequent items that have very sparse history interactions, which makes the training of CTR models non-trivial.
\item \textbf{Label Noise.} 
Besides sparsity, the collected user-item interactions are also \textit{noisy} by two reasons.
On the one hand, there exist spurious interactions derived from miss clicks or simple curiosity rather than users' true interests.
On the other hand, there are items that match one user's potential interests but are not interacted due to underexposure. However, the random negative sampling process may judge them as not interesting to the user. 
\end{itemize}

\subsection{Outline of MISS}
The label sparsity and label noise problems deteriorate quickly with the growth of data volume and feature size due to the Matthew Effect. 
In other words, popular items occupy more and more exposure chances and accumulate richer features and more supervision signals. On the contrary, unpopular items are less likely to be seen by users and suffer from the increasing lack of features and supervision.
Without sufficient and correct supervision, it is difficult for CTR models to make accurate predictions.
To this end, we propose a novel MISS framework to improve representation learning by means of SSL, as shown in Figure \ref{fig:framework}. Our MISS framework is model-agnostic and can be outlined with three major contributions: 
\begin{itemize}[leftmargin=*]
\item \textbf{Interest Augmentation.}
Two interest representation extractors are proposed to explore both point-wise and union-wise interest representations while considering inter-item and intra-item correlations. After that, the extracted interest representations are further augmented in consideration of the short-range and long-range dependencies, which makes them more robust to label noise.
\item \textbf{Contrastive Learning.}
Contrastive learning is imposed on the multi-interest representations, which effectively transforms the interest-level correlations into extra supervision signals to alleviate the lack of supervision caused by label sparsity.
\item \textbf{Model Compatibility.}
Under a multi-task learning framework, MISS flexibly combines the SSL component with any CTR prediction model in a plug-and-play manner,
which achieves both significant performance boosts and excellent compatibility with little handcrafted model configurations.
\end{itemize}

\section{Framework}\label{sec:framework}
\begin{figure*}[ht]
\centering
\setlength{\abovecaptionskip}{-0.2cm}
\includegraphics[width=5.4in]{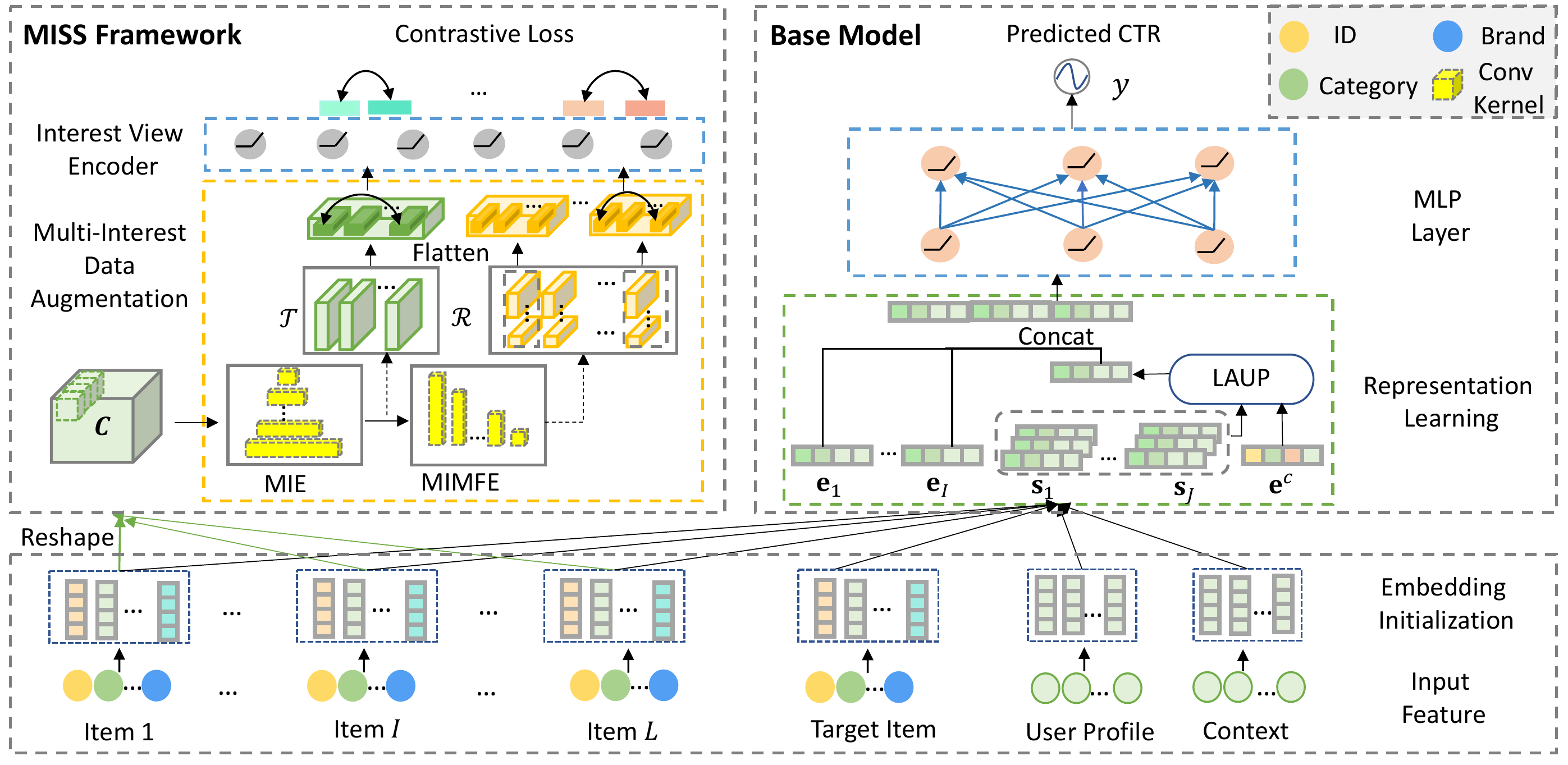}
\caption{Overview of our proposed MISS framework. The right part illustrates a typical deep CTR model, and the left part is the proposed multi-interest self-supervised learning component applied to it.} 
\label{fig:framework}
\vspace{-5mm}
\end{figure*}

In this section, we present the technical details of the proposed MISS framework.
As illustrated in the right part of Figure \ref{fig:framework}, a typical deep CTR model DIN \cite{zhou2018deep} is given as the default base model according to the experimental results in Table \ref{tab:compatibility},
based on which we explain in detail how the MISS framework is applied in a plug-and-play mechanism.
For other advanced deep CTR models, a compatibility analysis is also provided later in the experiments.

\subsection{The Base Model}
The base model consists of embedding initialization, representation learning, and CTR prediction components.

\subsubsection{Embedding Initialization}
In CTR prediction, input data samples are usually represented as high-dimensional sparse feature vectors as in Equation (\ref{equ:fea_vec}).
To facilitate follow-up calculations, feature vectors are first transformed into dense real-valued embedding vectors.
For the $I$ one-hot categorical features, $f_i$ is embedded into $\textbf{e}_i$ by looking up the field-wise embedding table.
While a sequential feature $\textbf{s}_j$ is represented as a list of embedding vectors.
By gathering all categorical and sequential feature embeddings, a sample is represented as a set of embedding vectors
\begin{equation}\label{equ:embed}
\setlength{\abovedisplayskip}{3pt}
\setlength{\belowdisplayskip}{3pt}
 \textbf{E} = \{\textbf{e}_1, \cdots, \textbf{e}_I, \textbf{e}_{1,1}, \cdots, \textbf{e}_{1,L}, \cdots, \textbf{e}_{J,1}, \cdots, \textbf{e}_{J,L}\},
\end{equation}
where $L$ is the sequence length.

 
\subsubsection{Representation Learning}
The behavior sequence length $L$ differs from user to user, thus the number of embeddings in $\textbf{E}$ also varies. 
A naive solution is to transform all sequential features into the same length with truncation and padding. However, truncation brings information loss and padding increases redundancy.
To handle this problem, researcher generally resort to different pooling techniques to aggregate the embedding sequences, which include max pooling, mean pooling, sum pooling, and the advanced attention-based pooling.
In this paper, we adopt the local activation unit based pooling, which was proposed in DIN \cite{zhou2018deep}, as it learns to assign an adaptive weight to each feature embedding according to the target item. For all $J$ sequential features with length $L$, the embedding vectors are aggregated into the sample representation as:
\begin{equation}\label{equ:embed_attention}
\setlength{\abovedisplayskip}{3pt}
\setlength{\belowdisplayskip}{3pt}
\textbf{X} = [\textbf{e}_1, \cdots, \textbf{e}_I, \text{LAUP}(\{\textbf{s}_1, \cdots, \textbf{s}_J\}, \textbf{e}^c)],
\end{equation}
where $\textbf{e}^c$ is the embedding of the candidate item, and $\text{LAUP}(\cdot, \cdot)$ is the pooling net based on the local activation unit. 
For space limitation, the technical details of $\text{LAUP}(\cdot, \cdot)$ is omitted here, interested readers can refer to \cite{zhou2018deep}.
Thus a fixed-length representation $\textbf{X}$ is obtained for each sample $x$.


\subsubsection{CTR Prediction}
Based on the integrated feature representation $\textbf{X}$, a Multi-Layer Perceptron (MLP) further learns the advanced feature interactions. Suppose a $D$-layer MLP is adopted, each layer works as
\begin{equation}\label{equ:mlp}
\setlength{\abovedisplayskip}{3pt}
\setlength{\belowdisplayskip}{3pt}
\textbf{a}^{(d)} = \sigma(\textbf{W}^{(d)}\textbf{a}^{(d-1)}+\textbf{o}^{(d)}),
\end{equation}
where $\textbf{a}^{(d-1)}$ is the output of the previous layer, $\sigma$ is the activation function, $\textbf{W}^{(d)}$ and $\textbf{o}^{(d)}$ are the weight matrix and bias vector respectively. 
We set $\textbf{a}^{(0)} = \textbf{X}$ for the first layer.
Finally, a prediction layer is devised to predict the CTR score
\begin{equation}\label{equ:sigmoid_predict}
\setlength{\abovedisplayskip}{3pt}
\setlength{\belowdisplayskip}{3pt}
\hat{y} = \text{Sigmoid}(\textbf{W}^{(D+1)}\textbf{a}^{(D)}+\textbf{o}^{(D+1)}),
\end{equation}
where $\text{Sigmoid}(\cdot)$ is the sigmoid activation function, and $\hat{y}$ is the predicted CTR score.
Finally, the batch-wise Logloss objective function is adopted to evaluate the predicted CTR score $\hat{y}$:
\begin{equation}
\setlength{\abovedisplayskip}{3pt}
\setlength{\belowdisplayskip}{3pt}
\mathcal{L}_{ll} = -\frac{1}{| \mathcal{B}|}\sum_{(x, y) \in  \mathcal{B} }{y\log\hat{y}+(1-y)\log(1-\hat{y})},
\end{equation}
where $\mathcal{B}$ is a batch of training samples, $(x, y)$ is the pair of sample and label in the batch, and $\mathcal{L}_{ll}$ is an instantiation of Equation (\ref{equ:obj}).

\subsection{The MISS Framework}
The base model estimates the CTR score through embedding initialization, representation learning, and CTR prediction, as illustrated in the right part of Figure \ref{fig:framework}. 
Based on that, the proposed MISS framework further enhances feature embeddings with SSL by multi-interest augmentation, interest view encoding, and contrative learning, as shown in the left part of Figure \ref{fig:framework}.
In this subsection, we explain the MISS components step-by-step.

\subsubsection{Sample-Level Data Augmentation}
Data augmentation is the first step of our proposed MISS framework, based on which the contrastive learning is implemented. 
Given a batch of training samples $\mathcal{B} = \{x_1, x_2, \cdots, x_{|\mathcal{B}|}\}$, existing SSL-based models all adopt \textsl{\textbf{sample-level}} data augmentation methods. 
For each sample $x$, two different views are first obtained through augmentation as:
\begin{equation}\label{equ:sample_aug}
\setlength{\abovedisplayskip}{3pt}
\setlength{\belowdisplayskip}{3pt}
\langle\textbf{h}^1, \textbf{h}^2\rangle = \text{Aug}^{s}(\textbf{x}),
\end{equation}
where $\text{Aug}^{s}(\cdot)$ is the sample-level augmentation function, and $\langle\textbf{h}^1, \textbf{h}^2\rangle$ is the pair of generated views.

After data augmentation, encoder functions are further used to extract high-level semantic representations from $\textbf{h}^1$ and $\textbf{h}^2$, based on which a contrastive loss is used to make use of the self-supervision signals.
However, due to the multi-interest characteristic of user behavior sequences, \textsl{\textbf{sample-level}} data augmentation may inevitably introduce noise. The reason is that the augmented $\textbf{h}^1$ and $\textbf{h}^2$ may be derived from different interests even if they are obtained from the same $\textbf{x}$. 
To solve this problem, we put forward an \textsl{\textbf{interest-level}} SSL framework, i.e., MISS, which augments the training data at the interest level in an end-to-end fashion.

\subsubsection{Multi-Interest Data Augmentation}
In consideration of the multi-interest characteristic of user behaviors, our MISS framework implements SSL within each sample at both the interest level and the feature level.
Therefore, not only the semantics provided by each training sample can be enriched, but also the modeling and utilization of long behavior sequences get promoted. 
To achieve these targets, we design a novel multi-interest extractor for data augmentation purpose.

To augment user behavior data at the interest level, 
the multiple interest representations of each user should first be extracted. An intuitive augmentation method is to directly divide the user behavior sequences according to item categories.
However, item categories are usually defined in coarse granularities and are not always available in data. Therefore, we propose a CNN-based multi-interest extractor which transforms the sample feature $\textbf{x}$ into a group of implicit interest representations 
\begin{equation}\label{equ:mie}
\setlength{\abovedisplayskip}{3pt}
\setlength{\belowdisplayskip}{3pt}
\mathcal{T} = \text{MIE}(\textbf{x}) = \{\textbf{t}_{1}, \textbf{t}_{2}, \cdots, \textbf{t}_{|\mathcal{T}|}\},
\end{equation}
where $\text{MIE}(\cdot)$ is the multi-interest extractor network, $\mathcal{T}$ is the output user interest representation sequence, and $\textbf{t}_{k}$ is the $k$-th interest representation extracted from $\textbf{x}$.
Moreover, for a fine-grained understanding and utilization of interest semantics, another CNN-based feature augmentation component is further designed to augment each interest representation at the feature level
\begin{equation}\label{equ:mief}
\setlength{\abovedisplayskip}{3pt}
\setlength{\belowdisplayskip}{3pt}
\begin{aligned}
\mathcal{R} &= \text{MIMFE}(\textbf{x}) \\
            &= \{\{\textbf{r}_{1,1}, \cdots, \textbf{r}_{1,\Omega}\}, \ldots, 
            \{\textbf{r}_{|\mathcal{T}|,1}, \cdots, \textbf{r}_{|\mathcal{T}|,\Omega}\}\}
\end{aligned}
\end{equation}
where $\text{MIMFE}(\cdot)$ is the multi-interest multi-feature extractor network that extracts fine-grained representations for each user interest, and  $\Omega$ is the number of feature representations for each interest.
After that, an augmentation function is applied to $\mathcal{T}$ to obtain interest-level augmented views as:
\begin{equation}\label{equ:select_t}
\setlength{\abovedisplayskip}{3pt}
\setlength{\belowdisplayskip}{3pt}
\mathcal{H}^{i} = \text{Aug}^{i}(\mathcal{T}) = \{\langle\textbf{h}_1^{i,1}, \textbf{h}_1^{i,2}\rangle, \cdots, \langle\textbf{h}_P^{i,1}, \textbf{h}_P^{i,2}\rangle\}, \\ 
\end{equation}
where $\text{Aug}^{i}(\cdot)$ is the interest-level augmentation function, $\langle\textbf{h}_p^{i,1}, \textbf{h}_p^{i,2}\rangle$ is a pair of generated views for sample $x$, and $P$ is the number of generated view pairs.
Similarly, an augmentation function is also applied to $\mathcal{R}$ for a further fine-grained augmentation as:
\begin{equation}\label{equ:select_r}
\setlength{\abovedisplayskip}{3pt}
\setlength{\belowdisplayskip}{3pt}
\begin{aligned}
\mathcal{H}^{if} &= \text{Aug}^{if}(\mathcal{R}) \\ 
                 &= \{\langle\textbf{h}_1^{if,1}, \textbf{h}_1^{if,2}\rangle, \cdots, \langle\textbf{h}_Q^{if,1}, \textbf{h}_Q^{if,2}\rangle\}, \\ 
\end{aligned}
\end{equation}
where $\text{Aug}^{if}(\cdot)$ is the feature-level augmentation function, $\langle\textbf{h}_q^{if,1}, \textbf{h}_q^{if,2}\rangle$ is a pair of views, and $Q$ is the number of generated pairs.
In this section, we focus on the principled explanation of our MISS framework, while the technical details of $\text{MIE}(\cdot)$, $\text{MIMFE}(\cdot)$, $\text{Aug}^{i}(\cdot)$, and $\text{Aug}^{if}(\cdot)$ are presented later in Section \ref{sec:methodology}.



\subsubsection{Interest View Encoder}
With the extractor networks and augmentation functions, two sequences of augmented view pairs are obtained, i.e., $\mathcal{H}^{i}$ and $\mathcal{H}^{if}$, where the user interest representations are augmented at different granularities.
Based on the sequence $\mathcal{H}^{i}$, an encoder is adopted to explore high-order abstractions:
\begin{equation}\label{equ:encode_1}
\setlength{\abovedisplayskip}{3pt}
\setlength{\belowdisplayskip}{3pt}
\mathcal{Z}^i = \text{Enc}^i(\mathcal{H}^{i}) = \{\langle\textbf{z}_1^{i,1}, \textbf{z}_1^{i,2}\rangle, \cdots, \langle\textbf{z}_P^{i,1}, \textbf{z}_P^{i,2}\rangle\}, 
\end{equation}
where $\text{Enc}^i(\cdot)$ is the encoder network that transforms each interet view representation $\textbf{h}_p^{i,1}$ (or $\textbf{h}_p^{i,2}$) into high-order representation $\textbf{z}_p^{i,1}$ (or $\textbf{z}_p^{i,2}$).
Similarly, an encoder $\text{Enc}^{if}(\cdot)$ is also applied to $\mathcal{H}^{if}$:
\begin{equation}\label{equ:encode_2}
\setlength{\abovedisplayskip}{3pt}
\setlength{\belowdisplayskip}{3pt}
\begin{aligned}
\mathcal{Z}^{if} &= \text{Enc}^{if}(\mathcal{H}^{if}) \\ 
                 &= \{\langle\textbf{z}_1^{if,1}, \textbf{z}_1^{if,2}\rangle, \cdots, \langle\textbf{z}_Q^{if,1}, \textbf{z}_Q^{if,2}\rangle\}. \\
\end{aligned}
\end{equation}
As we mainly focus on the extraction and utilization of self-supervised signals, two simple MLPs are used to implement $\text{Enc}^i(\cdot)$ and $\text{Enc}^{if}(\cdot)$.
However, other advanced networks are also applicable, such as Transformer in~\cite{sun2019bert4rec,xie2020contrastive}, and we leave the exploration of other encoder structures to future works.
\subsubsection{Contrastive Loss}
Having obtained the high-level semantics for each augmented view, the contrastive losses can finally be applied to exploit the self-supervision signals. 
Following SimCLR~\cite{chen2020simple}, we use the InfoNCE contrastive loss \cite{oord2018representation} which attempts to maximize the similarity of positive pairs of views and minimize the agreement of negative pairs of views.
As a result, similar interests can thus have similar representations (defined as \textit{alignment}) and sufficient information are kept to distinguish different interests (defined as \textit{uniformity}).
Both the \textit{alignment} and \textit{uniformity} properties are necessary and important for a good SSL system, as proved in~\cite{Wang2020Understanding}.
Formally, taking $\langle\textbf{z}_p^{i,1}, \textbf{z}_p^{i,2}\rangle$ from the same interest as positive pairs while $\langle\textbf{z}_p^{i,1}, \textbf{z}_p^{'i,2}\rangle$ and $\langle\textbf{z}_p^{'i,1}, \textbf{z}_p^{i,2}\rangle$ from different samples as negative pairs, the InfoNCE loss for learning the \textsl{\textbf{interest-level correlation}} is formulated as:
\setlength{\abovedisplayskip}{3pt}
\setlength{\belowdisplayskip}{3pt}
\begin{align}\label{equ:self-loss-1}
    \mathcal{L}_{ssl} = -\frac{1}{|\mathcal{B}| P}\sum_{x \in \mathcal{B}}{\sum_{1 \le p \le P}{\log \frac{\exp(s(\textbf{z}_p^{i,1}, \textbf{z}_p^{i,2})/\tau)}{\sum_{x' \in \mathcal{B}} \exp(s(\textbf{z}_p^{i,1}, \textbf{z}_p^{'i,2})/\tau)}}},
\end{align}
where $s(\cdot, \cdot)$ is cosine similarity function, $\tau$ is the softmax temperature parameter, and $\exp(\cdot)$ is the exponential function.
Similarly, the InfoNCE loss for learning the \textsl{\textbf{feature-level correlation}} can be formulated as:
\setlength{\abovedisplayskip}{3pt}
\setlength{\belowdisplayskip}{3pt}
\begin{align}\label{equ:self-loss-2}
    \mathcal{L}_{ssl}^{'}=  -\frac{1}{|\mathcal{B}| Q}\sum_{x \in \mathcal{B}}{\sum_{1 \le q \le Q}{\log \frac{\exp(s(\textbf{z}_q^{if,1}, \textbf{z}_q^{if,2})/\tau)}{\sum_{x' \in \mathcal{B}} \exp(s(\textbf{z}_q^{if,1}, \textbf{z}_q^{'if,2})/\tau)}}}.
\end{align}
\subsection{Multi-task Learning}
To better integrate the MISS framework with the CTR prediction component, a multi-task learning strategy is adopted to jointly optimize the auxiliary SSL losses and the main prediction loss in an end-to-end manner. Thus the final loss function is formulated as:
\begin{align}\label{equ:final loss}
\setlength{\abovedisplayskip}{3pt}
\setlength{\belowdisplayskip}{3pt}
     \mathcal{L} = \mathcal{L}_{ll} + \alpha_1 \cdot \mathcal{L}_{ssl} + \alpha_2 \cdot \mathcal{L}_{ssl}^{'},
\end{align}
where $\alpha_1$ and $\alpha_2$ are the hyper-parameters to control the strength of SSL losses.
In experiments, the two-stage pre-training learning strategy is also tried, where the model is first trained with the auxiliary SSL losses, then fine-tuned by the main prediction loss.

\section{Multi-Interest Data Augmentation}\label{sec:methodology}
\begin{figure}[t]
\centering
\setlength{\abovecaptionskip}{-0.2cm}
\noindent\makebox[0.45\textwidth][c]{\includegraphics[scale=0.3]{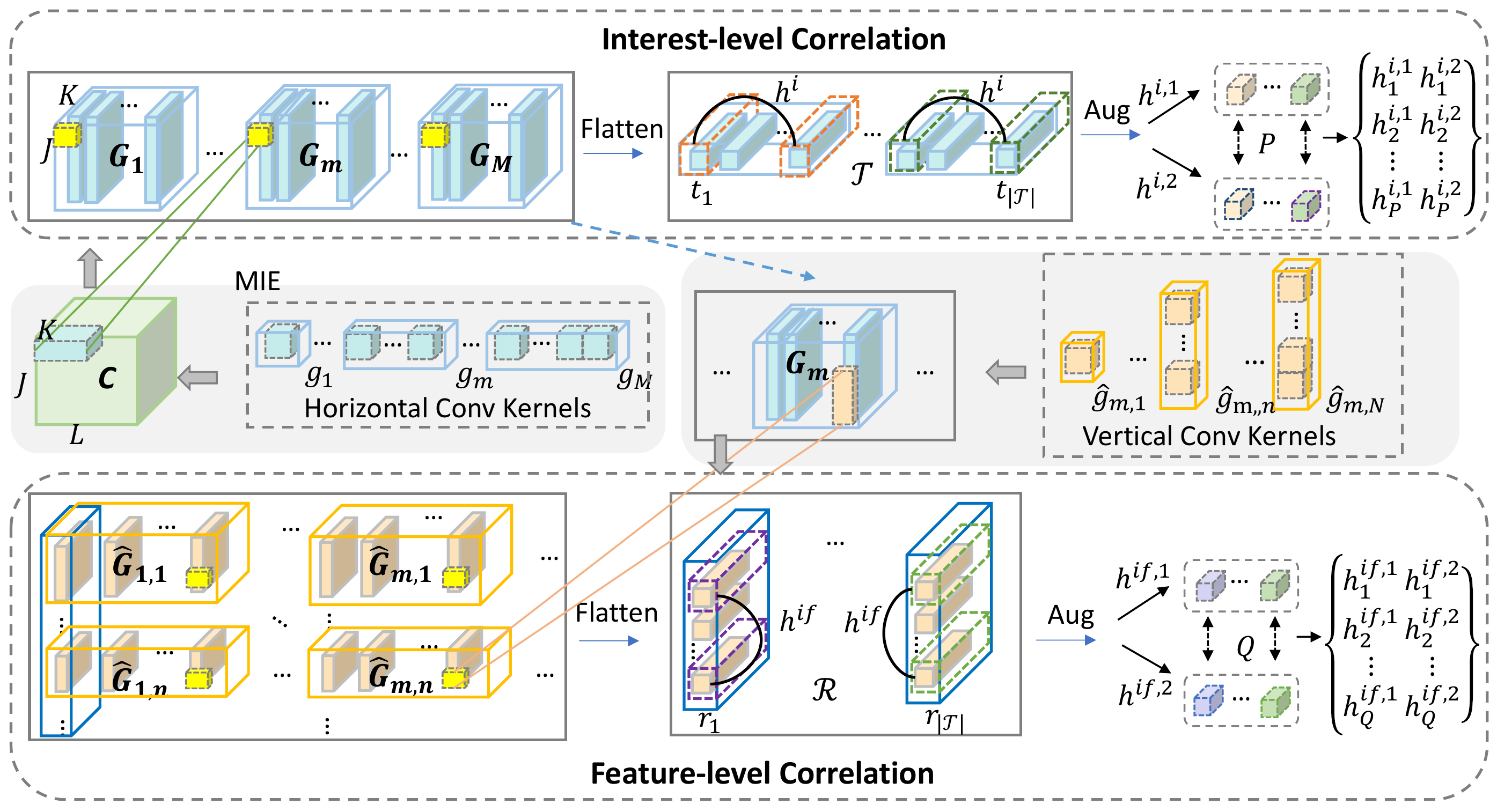}}
\caption{Multi-interest data augmentation.}
\label{fig:multi-interest}
\vspace{-5mm}
\end{figure}
As mentioned in Section \ref{sec:framework}, a multi-interest extractor network $\text{MIE}(\cdot)$ and an interest-level feature extractor network $\text{MIMFE}(\cdot)$ are used for interest representation learning at different granularities, based on which the augmentation functions $\text{Aug}^i(\cdot)$ and $\text{Aug}^{if}(\cdot)$ are further applied. In this section, we describe the technical details of these extractors and augmentation functions.
\subsection{Multi-Interest Extractor}
The multi-interest extractor network $\text{MIE}(\cdot)$ aims to discover the potential interests from user behavior sequences. However, it is hard to achieve this goal as the number of interests varies from user to user.
Moreover, the sequential pattern of the same interest is also dynamic in terms of both different users and different time. 
Therefore, we propose an intuitive multi-interest extractor based on a closeness assumption, i.e., user behaviors derived from the same interest are more likely to be closely located within a sequence.

Based on the closeness assumption, we adopt CNN to extract hidden interest representations due to its effectiveness in capturing the local correlations~\cite{krizhevsky2012imagenet,he2016deep}.
Other sequence representation learning models like RNNs or self-attention are also applicable, however, they fail to extract effective interest representation pairs for comparison. We will verify this point later in the experiments.
After padding, all $J$ sequential features share the same length $L$, and the embeddings in $\textbf{E}$ can be re-organized into a 3D tensor as:
\begin{equation}\label{eq3}
\setlength{\abovedisplayskip}{3pt}
\setlength{\belowdisplayskip}{3pt}
\textit{\textbf{C}} = \begin{bmatrix}
\textbf{e}_{1,1}, \textbf{e}_{1,2}, \cdots, \textbf{e}_{1,L}\\
\textbf{e}_{2,1}, \textbf{e}_{2,2}, \cdots, \textbf{e}_{2,L}\\
\vdots \\
\textbf{e}_{J,1}, \textbf{e}_{J,2}, \cdots, \textbf{e}_{J,L}\\
\end{bmatrix}
\end{equation}
where $\textit{\textbf{C}} \in \mathbb{R}^{J \times L \times K}$, and $K$ is dimension of each embedding vectors $\textbf{e}_{j,l}$. 
Hidden interests are extracted through horizontal convolutions along the time axis of $\textit{\textbf{C}}$. 
Take the click sequence in Figure \ref{fig:interest} as an example, one  convolution kernel may aggregate the feature embeddings of the paint board ($\textit{\textbf{C}}^{:,1,:}$), the brush ($\textit{\textbf{C}}^{:,2,:}$), and the palette ($\textit{\textbf{C}}^{:,3,:}$) into the interest in painting tools, while another kernel may take the embeddings of the notebook ($\textit{\textbf{C}}^{:,4,:}$) alone as the interest in electronic products.


Specifically, a horizontal convolution layer with $M$ branches of kernels are adopted, as shown in left middle part of Figure \ref{fig:multi-interest}.
Denote $g_m \in \mathbb{R}^{1 \times m \times 1}$ as a kernel with width $m \in [1, M]$, where both the kernel height and channel number are set to 1. For simplification, only one kernel is used in each branch. 
Thus $M$ kernels with different widths are used, which simultaneously capture the point-wise ($m=1$) and union-wise ($m>1$) interest representations.
Each $g_m$ slides on $\textit{\textbf{C}}$ from left to right, and the convolution operation at the $j$-th row, the $l$-th to the $(l+m-1)$-th column, and the $k$-th channel in $\textit{\textbf{C}}$ is formulated as:
\begin{equation}\label{eq4}
\setlength{\abovedisplayskip}{3pt}
\setlength{\belowdisplayskip}{3pt}
G^{j,l,k}_{m} = \text{ReLU}(\textit{\textbf{C}}^{j,l:l+m-1,k} \circ g_m),
\end{equation} 
where $\text{ReLU}(\cdot)$ is the ReLU activation function, $\circ$ denotes the convolution operation, $\textit{\textbf{C}}^{j,l:l+m-1,k}$ represents the sliced sub-tensor from $\textit{\textbf{C}}$, and $l$ is restricted as $1\leq l\leq (L-m+1)$. After convolution, the final output tensor of $g_m$ is denoted as $\textit{\textbf{G}}_{m} \in \mathbb{R}^{J \times (L-m+1) \times K}$, which is a combination of $(L-m+1)$ interest representations.
Take all $M$ filters as a whole, the resulting $|\mathcal{T}| = \sum_{1 \leq m \leq M}{(L-m+1)}$ interest representations together make up the interest sequence
\begin{equation}\label{equ:mie_detail}
\setlength{\abovedisplayskip}{3pt}
\setlength{\belowdisplayskip}{3pt}
\begin{aligned}
\mathcal{T} &= \text{MIE}(\textbf{x}) \\ 
            &= \{ \ldots, \text{Flat}(\textit{\textbf{G}}_m^{:,l,:}), \ldots \} \\
            &= \{\textbf{t}_{1}, \textbf{t}_{2}, \cdots, \textbf{t}_{|\mathcal{T}|}\},
\end{aligned}
\end{equation}
where $\text{Flat}(\cdot)$ is the flatten function that transforms each interest representation $\textit{\textbf{G}}_m^{:,l,:} \in \mathbb{R}^{J \times K}$ into a vector $\textbf{t} \in \mathbb{R}^{JK}$, as shown in the upper part of Figure \ref{fig:multi-interest}.
\subsection{Interest-Level Augmentation}
The multi-interest extractor learns different user interest representations based on the closeness assumption. However, the closeness assumption not only holds at the behavior level, but also applies to the extracted interest representations. 
In other words, the more adjacent two interest representations are located on the time line, the more likely they represent the same hidden interest.
Therefore, we randomly select a pair of representations as two different views of the same interest from those with the same filter $g_m$ from $\mathcal{T}$
\begin{equation}\label{equ:select_t_detail}
\setlength{\abovedisplayskip}{3pt}
\setlength{\belowdisplayskip}{3pt}
\begin{aligned}
\mathcal{H}^{i} &= \text{Aug}^{i}(\mathcal{T}) \\ 
                &= \{\ldots, \text{RS}^{i}(\textit{\textbf{G}}_{m}), \ldots \} \\
                &= \{\ldots, \langle\textbf{t}_l, \textbf{t}_{l+h}\rangle, \ldots\} \\
                &= \{\langle\textbf{h}_1^{i,1}, \textbf{h}_1^{i,2}\rangle, \cdots, \langle\textbf{h}_P^{i,1}, \textbf{h}_P^{i,2}\rangle\}, \\ 
\end{aligned}
\end{equation}
where $\text{RS}^{i}(\cdot)$ randomly selects two representations derived from the same convolution filter with a given distance $h \in [1, H]$. By repeating the select function $\text{RS}^{i}(\cdot)$ for $P$ times, the sequence of augmented interest view pairs $\mathcal{H}^{i}$ is obtained.
Here we use different $h$ values to cover both short-range and long-range interest dependencies, and a maximum distance $H$ is pre-defined to prevent overlong dependencies.
Note that, we assume a uniform distribution of interest dependency distance $h$. However, other complex distributions (e.g., Gaussian distribution) are also applicable, and we leave them to future works. 

\subsection{Fine-Grained Interest Extractor}
The multi-interest extractor $\text{MIE}(\cdot)$ only explores the inter-item correlations of different interests along the time line, while the intra-item relationship between all $J$ features within each interest is ignored.
For example, given the resulting interest representations $\textit{\textbf{G}}_{m} \in \mathbb{R}^{J \times (L-m+1) \times K}$, the interest in daily supplies (say $\textit{\textbf{G}}_{m}^{:,l,:}$) is sensitive to price, while the interest in shoes (say $\textit{\textbf{G}}_{m}^{:,l',:}$) is affected by both price and category.
To deal with this issue, vertical convolution operators are further applied to each resulting at the feature level for fine-grained augmentation.

As shown in the right middle part of Figure \ref{fig:multi-interest}, $N$ branches of vertical kernels $\hat{g}_{m,n} \in \mathbb{R}^{n \times 1 \times 1}$ are adopted to learn intra-item correlations, where $n \in [1, N]$ is the kernel height.
As $n$ varies from $1$ to $N$, both single and collective feature representations are captured.
Each kernel $\hat{g}_{m,n}$ interacts with the sliced tensor $\textit{\textbf{G}}_m^{j:j+n-1,l,k}$ from left to right, which yields the result $\hat{G}^{j,l,k}_{m,n}$:
\begin{equation}\label{eq5}
\setlength{\abovedisplayskip}{3pt}
\setlength{\belowdisplayskip}{3pt}
\hat{G}^{j,l,k}_{m,n} =\text{ReLU}(\textit{\textbf{G}}_m^{j:j+n-1,l,k} \circ \hat{g}_{m,n}).
\end{equation}
Denote the final output via $\hat{g}_{m,n}$ on $\textit{\textbf{G}}_{m}$ as $\hat{\textit{\textbf{G}}}_{m,n} \in \mathbb{R}^{(J-n+1) \times (L-m+1) \times K}$, it can also be viewed as a combination of $(J-n+1)$ interest representations.
However, the interest representations are now enhanced at the feature level.
Through all $N$ kernels, $|\mathcal{T}|\Omega$ enhanced interest representations are obtained, where $\Omega = \sum_{1 \leq n \leq N}{(J-n+1)}$. Formally, 
\begin{equation}\label{equ:mief-detailed}
\setlength{\abovedisplayskip}{3pt}
\setlength{\belowdisplayskip}{3pt}
\begin{aligned}
\mathcal{R} &= \text{MIMFE}(\textbf{x}) \\
            &= \{\ldots, \{\ldots, \text{Flat}(\hat{\textit{\textbf{G}}}_{m,n}^{j,:,:}) , \ldots\}, \ldots\} \\
            &= \{\{\textbf{r}_{1,1}, \cdots, \textbf{r}_{1,\Omega}\}, \ \ldots,  \{\textbf{r}_{|\mathcal{T}|,1}, \cdots, \textbf{r}_{|\mathcal{T}|,\Omega}\}\}.
\end{aligned}
\end{equation}
\subsection{Fine-Grained Interest-Level Augmentation}
Because of the independence between each feature, a totally random select function $\text{RS}^{if}(\cdot)$ is applied to sample feature-level interest representation views from each $\hat{\textit{\textbf{G}}}_{m,n}$, as
\begin{equation}\label{equ:select_r_detail}
\setlength{\abovedisplayskip}{3pt}
\setlength{\belowdisplayskip}{3pt}
\begin{aligned}
\mathcal{H}^{if} &= \text{Aug}^{if}(\mathcal{R}) \\ 
                 &= \{\ldots, \text{RS}^{if}(\hat{\textit{\textbf{G}}}_{m,n}), \ldots \} \\
                 &= \{\ldots, \langle\textbf{r}_j, \textbf{r}_{j'}\rangle, \ldots \} \\
                 &= \{\langle\textbf{h}_1^{if,1}, \textbf{h}_1^{if,2}\rangle, \cdots, \langle\textbf{h}_Q^{if,1}, \textbf{h}_Q^{if,2}\rangle\}, \\ 
\end{aligned}
\end{equation}
which is repeated for $Q$ times.
\subsection{Complexity Analysis}
For the $M$ branches of horizontal convolution kernels, i.e., $g_m \in \mathbb{R}^{1 \times m \times 1}$, there are $m$ learnable parameters.
As $m$ ranges from $1$ to $M$, the total number of learnable parameters is $\sum_{1 \leq m \leq M}{m} = \frac{M(M-1)}{2}$.
Similarly, the total number of  the $N$ branches of vertical convolution kernels is $\sum_{1 \leq n \leq N}{n} = \frac{N(N-1)}{2}$.
After data augmentation, two encoders are used for high-order abstraction.
For simplicity, two MLP encoders are used.
Thus the numbers of introduced parameters are $J \times K \times H^{i}_1 + \sum_{2 \leq d \leq D^{'}}{H^{i}_{d-1} \times H^{i}_{d}}$ for encoder $\text{Enc}^i(\cdot)$, and $K \times H^{if}_1 + \sum_{2 \leq d \leq D^{'}}{H^{if}_{d-1} \times H^{if}_{d}}$  for encoder $\text{Enc}^{if}(\cdot)$, where $D^{'}$ is the layer depth and $H^{*}_d$ is the layer size of the $d$-th layer.
All together, the total number of parameters brought by the multi-interest data augmentation in MISS is $\frac{M(M-1) + N(N-1)}{2} + J \times K \times H^{i}_1 + K \times H^{if}_1 + \sum_{2 \leq d \leq D^{'}}{H^{i}_{d-1} \times H^{i}_{d} + H^{if}_{d-1} \times H^{if}_{d}}$.
On the whole, $M$ and $N$ are very small values and the MLP parameters are also negligible compared to the embedding matrices, which makes the space complexity of MISS acceptable.

\section{Experiments}
\subsection{Experiment Setup}\label{ExperimentSetup}
\subsubsection{Datasets}
We evaluate the effectiveness of our proposed model on three large-scale datasets, i.e., \textit{Amazon-Cds}, \textit{Amazon-Books}, and \textit{Alipay}. All three of them are real-world datasets described as follows:
\begin{itemize}[leftmargin=*]
    \item \textbf{Amazon Dataset}\footnote{http://jmcauley.ucsd.edu/data/amazon/}: 
    The Amazon dataset collects user review data from one of the largest e-commerce website in the world, i.e., amazon.com.
    The crawled reviews have a time span from May 1996 to July 2014.
    The dataset can be divided into many subsets according to the various product categories, such as Amazon-Electronics, Amazon-Cds, and Amazon-Books.
    In this paper, we pick the Amazon-Cds and Amazon-Books subsets for experiments.
	\item \textbf{Alipay}\footnote{https://tianchi.aliyun.com/dataset/dataDetail?dataId=53}: 
	The Alipay dataset is provided in the IJCAI-16 contest which is collected from the Tmall.com website, the Taobao.com website, and the Alipay App.
	It contains user behavior logs between July 1st to November 30th in 2015.
	Each log contains multiple feature fields, including user ID, item ID, seller, category, online action type, and timestamp.
	We take the click behaviors as users' interaction records to construct the user behavior sequences.
\end{itemize}
Table \ref{tab:dataset} presents the detailed statistics of the three datasets. As we can see, the datasets are different from each other in many aspects including feature number, field number, and interaction sparsity.
\begin{table}[ht]
\vspace{-2mm}
\setlength{\abovecaptionskip}{-0.1cm}
\setlength{\belowcaptionskip}{-0.1cm}
 \caption{Dataset statistics.}
 \centering
 	\setlength{\tabcolsep}{1mm}
 \begin{tabular}{c|c|c|c|c|c}
 \hline
 Dataset 	 & \#Users & \#Items & \#Instances & \#Features & \#Fields  \\
 \hline
 Amazon-Cds & 75,258 & 64,443 & 150,516 & 140,167 & 5  \\
 Amazon-Books & 158,650 & 128,939 & 317,300 & 288,577 & 5   \\
 Alipay & 326,577 & 451,631 & 653,154 & 788,166 & 7 \\ 
 \hline
\end{tabular}
\label{tab:dataset}
\vspace{-3mm}
\end{table}
\subsubsection{Data Processing}
To ensure the data quality, we filter out infrequent users and items with fewer than 5 interactions in the Amazon-Cds dataset. While the threshold value is 10 in the Amazon-Books and Alipay datasets. 
For all datasets, we aggregate each user's interaction records and
sort them by the action timestamps in chronological order. For evaluation purpose, we adopt the data split strategy in \cite{ren2019lifelong,qin2020user}.
Specifically, suppose a user has $L$ historical behaviors sorted by time, the behavior sequence $[1, L-3]$ is used for training and predicts whether she/he will interact with the $(L-2)$-th item. 
Similarly, behavior sequence $[1, L-2]$ is used to predict the $(L-1)$-th item in the validation set, while behavior sequence $[1, L-1]$ is used to predict the $L$-th item in the testing set.
Further, given a user, a non-interacted item is randomly selected as the negative sample.

\subsubsection{Baseline Models}
For a thorough verification of model effectiveness, the proposed MISS framework is compared with three groups of representative CTR prediction models: a) Feature interaction based models (LR \cite{lee2012estimating}, FM \cite{rendle2010factorization}, DeepFM \cite{guo2017deepfm}, IPNN \cite{qu2018product}, DCN~\cite{wang2017deep}, DCN-M~\cite{wang2021dcn}, xDeepFM \cite{lian2018xdeepfm}); b) User interest modeling based models (DIN \cite{zhou2018deep}, DIEN \cite{zhou2019deep}), SIM(soft)~\cite{Pi2020Search}, DMR~\cite{lyu2020deep}; c) GNN and Transformer based models (AutoInt+ \cite{song2019autoint}, FiGNN \cite{li2019fi}).


\subsubsection{Evaluation Metrics}
To quantitatively evaluate the model performances, two widely-used metrics are adopted, i.e., \textit{AUC} and \textit{Logloss} \cite{guo2017deepfm}, which are widely used evaluation metrics for CTR prediction task.

\subsubsection{Parameter Settings}
For a fair comparison, we set the embedding dimension of all models as 10, the batch size is fixed as 128, and the learning rate is selected from \{$10^{-1}$, $10^{-2}$, $10^{-3}$, $10^{-4}$\}.
The deep layers for all models are set as \{40, 40, 40, 1\}. 
The Adam optimizer \cite{kingma2014adam} is chosen for model optimization.
In addition to the above hyper-parameters for all models, we set the layers for interest encoder and feature encoder as \{20, 20\} and \{10, 10\}. 
For simplicity, we set $\alpha_1 = \alpha_2$, and search them and $\tau$ within the ranges of \{0.05, 0.1, 0.5, 1, 5\}.
For the branches of horizontal and vertical convolution kernels, $M$ is tuned from \{1, 2, 3, 4\}, and $N$ is tuned from \{1, 2\}.
The distance $H$ is tuned from \{1, 2, 3, 4\}.
We use the validation set for parameter tuning, while the final reported performances are obtained on the testing set.
Each experiment is repeated for 5 times to remove random noises, and the averaged results are reported.

\subsection{Performance Comparison}
\begin{table}[ht]
\vspace{-3mm}
\setlength{\abovecaptionskip}{-0.1cm}
\setlength{\belowcaptionskip}{-0.1cm}
\centering
\caption{The overall performances on all three datasets. The $\star$ mark indicates the statistical significance (p-value$<$0.05) of the comparison between MISS and the strongest baseline (underlined values).}
\setlength{\tabcolsep}{1mm}{
\small
\begin{tabular}{c|c|c|c|c|c|c}
\hline \hline
Dataset &
\multicolumn{2}{c|}{Amazon-Cds} & 
\multicolumn{2}{c|}{Amazon-Books} &
\multicolumn{2}{c}{Alipay} \\ \hline 
Model & AUC & Logloss &  AUC & Logloss  & AUC & Logloss \\\hline \hline
LR & 0.6918 & 0.6308 & 0.7350 & 0.5968 &  0.7848  &0.5578  \\
FM  & 0.7585 & 0.5851 & 0.7653  & 0.5745  & 0.8470 &0.4859  \\\hline
DeepFM & 0.8039 & 0.5369 & 0.8056 & 0.5310  & 0.8718  & 0.4464   \\
IPNN   & 0.8053 & 0.5364& 0.8051 &  0.5308  & 0.8823& 0.4299  \\
DCN    & 0.7994 & 0.5412& 0.7982 & 0.5390 & 0.8700  & 0.4494   \\
DCN-M & 0.8050  & 0.5363 &0.8070 & 0.5293   & 0.8757 &0.4403  \\
xDeepFM & 0.8034 & 0.5370 & 0.8028 & 0.5336 & 0.8777 & 0.4382 \\\hline
DIN  & 0.8055 & 0.5357 & 0.8074  & 0.5289  & 0.9098  & 0.3734    \\
DIEN & 0.7928 & 0.5479 & 0.8016 & 0.5352 & 0.9004 & 0.3950 \\
SIM(soft) & 0.7977  & 0.5437 & 0.7951  &  0.5430   & 0.9101  & 0.3729 \\
DMR & \underline{0.8115}  & \underline{0.5289} & \underline{0.8082}  &  \underline{0.5282}   & \underline{0.9148}  & \underline{0.3642} \\\hline
AutoInt+  & 0.8008 & 0.5398 & 0.8045  & 0.5317  & 0.8705  & 0.4479    \\
FiGNN  & 0.8012 & 0.5391 & 0.8006 & 0.5366 & 0.8716 & 0.4467   \\ \hline
MISS & $\textbf{0.8867}^{\star}$ & $\textbf{0.4357}^{\star}$ & $\textbf{0.9180}^{\star}$ & $\textbf{0.3730}^{\star}$ & $\textbf{0.9327}^{\star}$ & $\textbf{0.3295}^{\star}$ \\
\hline \hline
\end{tabular}}
\vspace{-3mm}
\label{tab:ctraccuracy}
\end{table}
In this section, we compare the performances of MISS with the state-of-the-art CTR prediction models.
Table \ref{tab:ctraccuracy} shows the experimental results of all compared models on all three datasets.
From Table \ref{tab:ctraccuracy}, we have the following observations:
\begin{itemize}[leftmargin=*]
    \item MISS consistently performs better than all baselines on all three datasets.
    More precisely, MISS significantly ($p-value < 0.05$) outperforms the strongest baselines by  \textbf{9.27\%}, \textbf{13.55\%} and \textbf{1.96\%} in terms of \textit{AUC} (\textit{17.62\%}, \textit{29.38\%} and \textit{9.53\%} in terms of \textit{Logloss}) on the Amazon-Cds, Amazon-Books, and Alipay datasets respectively. 
    The great improvements over baseline models verify the effectiveness of MISS for CTR prediction.
    By supplementing the CTR prediction task with self-supervised learning, MISS is capable of exploiting the latent correlation information with more supervision signals, while baseline models only utilize the observed user-item interactions as supervision signals.
    
    \item The improvements in the Amazon-Cds and Amazon-Books datasets are much more significant than in the Alipay dataset.
    A possible reason is that the time span of user behaviors in these two datasets (over ten years) is much longer than that in the third dataset (six months).
    As more diverse interests take place in the relatively longer time span, our proposed MISS obtains more significant improvements by considering the multi-interest characteristic of user behaviors.
    \item LR and FM perform the worst among all baselines, which indicates that shallow models are insufficient for CTR prediction. 
    By modeling high-order feature interactions with DNNs, DeepFM, IPNN, DCN, DCN-M and xDeepFM perform better than shallow models.
    DIN, DIEN, SIM(soft), and DMR achieve comparable performances with deep feature interaction models, which demonstrates the usefulness of user interest mining. 
    DMR achieves the best performances among all compared baselines. A possible reason is that it learns better representations by utilizing and integrating both user-item and item-item interactions in an attentive manner.
    AutoInt+ and FiGNN use self-attention or GNN for feature interaction modeling. Similar performances can be found compared with DeepFM and IPNN. It indicates that only using user-item interactions as supervision signals (as by existing deep CTR models)  cannot make a big difference on the model performances.
\end{itemize}

\subsection{Ablation Study}
To better understand the design rational of our proposed MISS, we conduct a series of ablation experiments and analysis in this section.

\begin{table}[ht]
\vspace{-5mm}
\setlength{\abovecaptionskip}{-0.1cm}
\setlength{\belowcaptionskip}{-0.1cm}
\caption{Compatibility analysis results.}
    \centering
\setlength{\tabcolsep}{1mm}{
\small
\begin{tabular}{c|c|c|c|c|c|c}
\hline \hline 
Dataset &
\multicolumn{2}{c|}{Amazon-Cds} & 
\multicolumn{2}{c|}{Amazon-Books} &
\multicolumn{2}{c}{Alipay} \\ \hline 
Model & AUC & Logloss &  AUC & Logloss  & AUC & Logloss \\\hline \hline
DIN  & 0.8055 & 0.5357 & 0.8074 & 0.5289 & 0.9098 & 0.3734  \\
DIN-MISS  & \textbf{0.8867} & \textbf{0.4357} & \textbf{0.9180} & \textbf{0.3730} & \textbf{0.9327} & \textbf{0.3295}  \\\hline
IPNN & 0.8053 & 0.5364 &0.8051 & 0.5308 & 0.8823 & 0.4299  \\
IPNN-MISS  & \textbf{0.8858} & \textbf{0.4368} & \textbf{0.9146} & \textbf{0.3778} & \textbf{0.9004}  & \textbf{0.4006}  \\ \hline
FiGNN & 0.8012 & 0.5391 & 0.8006 & 0.5366 & 0.8716 & 0.4467  \\
FiGNN-MISS  & \textbf{0.8828} & \textbf{0.4410} & \textbf{0.9170} & \textbf{0.3746} & \textbf{0.8947} & \textbf{0.4160} \\
\hline \hline
\end{tabular}}
\vspace{-3mm}
\label{tab:compatibility}
\end{table}

\subsubsection{Compatibility Analysis}
Compatibility is among the key factors that restrict one model's applications. 
To verify the compatibility of our proposed MISS framework, apart from the DIN backbone model described in the framework section, we also use it to improve the representation learning in another two representative CTR models, i.e., IPNN, and FiGNN. For a fair comparison, other parts of these models remain unchanged, and the enhanced models are named as DIN-MISS (the same model as MISS), IPNN-MISS, and FiGNN-MISS respectively.
We compare the original and enhanced models on the three datasets, and the experimental results are presented in Table \ref{tab:compatibility}.
As can be easily observed, all three enhanced models (DIN-MISS, IPNN-MISS, and FiGNN-MISS) significantly outperform their original models on all three datasets. 
It validates the compatibility of our embedding enhancement approach by demonstrating its effectiveness when combined with various popular CTR models.
The results show that MISS can be used as a general framework to improve the existing CTR models by supplementing self-supervised signals for embedding enhancement. 
\begin{table}[ht]
\vspace{-5mm}
\setlength{\abovecaptionskip}{-0.1cm}
\setlength{\belowcaptionskip}{-0.1cm}
\caption{Superiority analysis results.}
    \centering
\setlength{\tabcolsep}{1mm}{
\begin{tabular}{c|c|c|c|c|c|c}
\hline \hline 
Dataset &
\multicolumn{2}{c|}{Amazon-Cds} & 
\multicolumn{2}{c|}{Amazon-Books} &
\multicolumn{2}{c}{Alipay} \\ \hline 
Model & AUC & Logloss & AUC & Logloss  & AUC & Logloss \\\hline \hline
IPNN  & 0.8053 & 0.5364 & 0.8051 & 0.5308 & 0.8823 & 0.4299   \\
IPNN-Rule & 0.8101  & 0.5303 &0.8497 & 0.4788   & 0.8818 &0.4321 \\
\footnotesize{IPNN-IRSSL}  &0.8050 &0.5371 & 0.8065 & 0.5295 & 0.8821 & 0.4314   \\
\footnotesize{IPNN-S3Rec}  & 0.8065 & 0.5343 & 0.8073 & 0.5286 & 0.8826 & 0.4299   \\
\footnotesize{IPNN-CL4SRec}  & \underline{0.8372} & \underline{0.5057} &  \underline{0.8759} & \underline{0.4553} & \underline{0.8865} & \underline{0.4250}  \\
\footnotesize{IPNN-MISS} & \textbf{0.8858} & \textbf{0.4368}  & \textbf{0.9146} & \textbf{0.3778} & \textbf{0.9004}  & \textbf{0.4006} \\ \hline

DIN & 0.8055 & 0.5357 & 0.8074  & 0.5289 & 0.9098 & 0.3734 \\
DIN-Rule & 0.8068 & 0.5349 &0.8397 & 0.4925   & 0.9113 &0.3719  \\
\footnotesize{DIN-IRSSL}  & 0.8058 & 0.5352 & 0.8064 & 0.5295 & 0.9098 & 0.3744  \\
\footnotesize{DIN-S3Rec}  & 0.8073 & 0.5348 & 0.8076 & 0.5286 & 0.9100  & 0.3728 \\
\footnotesize{DIN-CL4SRec}  & \underline{0.8364} &  \underline{0.5082}  & \underline{0.8756} & \underline{0.4563} & \underline{0.9141} & \underline{0.3686}  \\
\footnotesize{DIN-MISS} & \textbf{0.8867} & \textbf{0.4357} & \textbf{0.9180} & \textbf{0.3730} & \textbf{0.9327} & \textbf{0.3295} \\
\hline \hline
\end{tabular}}
\label{tab:superioty}
\vspace{-3mm}
\end{table}
\subsubsection{Superiority Analysis}
To demonstrate the superiority of our proposed MISS framework, we compare it with state-of-the-art self-supervised learning models. Specifically, we apply MISS, IRSSL~\cite{yao2021selfsupervised}, S3Rec~\cite{zhou2020s3}, and CL4SRec~\cite{xie2020contrastive} to the IPNN, DIN, and FiGNN models for embedding enhancement purpose. 
Besides above SSL models, we also equip these base CTR models with a rule based model that segments the behavior sequence into several sub-sequences based on item categories and then conduct dropout on each sequence for SSL.
The resulting models are named in an ``A''-``B'' manner where ``A'' and ``B'' represent the base model and the SSL method respectively.
Notice that, we adopt the item feature mask strategy in IRSSL as it achieves better performances than feature dropout, and the sequence-segment correlation is adopted in S3Rec thanks to its best performances within the four data augmentation techniques.
Comparative experimental results of the original and enhanced models are shown in Table \ref{tab:superioty}. 
Due to the space limitation and similar trend of evaluation metrics, results of the FiGNN model are not presented.
From Table \ref{tab:superioty}, we have the following findings:
\begin{itemize}[leftmargin=*]
\item Our MISS model consistently performs the best regardless of the base models or datasets, which further verifies the superiority of our comparative learning strategies.

\item Rule based SSL model achieves much better performances than IRSSL on the Amazon-Books dataset, which verifies the effectiveness of interest-level contrastive learning for recommendation tasks.
However, comparable performances are achieved on the other two datasets.
The reason is that the item categories in different datasets are differently defined. 
In some cases, item categories indicate user interests well, but in other cases they do not.

\item Generally speaking, IPNN-IRSSL performs no better than IPNN and it is the same for DIN-IRSSL and DIN. The reason is that IRSSL only focuses on item features, thus loses efficacy when few item features are available.

\item IPNN-S3Rec and DIN-S3Rec perform slightly better than the original models, which supports the effectiveness of SSL at the behavior level. 
However, there is an obvious semantic difference between a random segment and the whole behavior sequence, hence the correlation learning is biased and limits its performances.

\item Models enhanced by the CL4SRec method achieve the second best performances. In CL4SRec, the majority of the behavior sequences remain unchanged after the item crop, mask, and reorder operations, which makes it more robust to random noises. In our MISS, however, a more flexible data augmentation method is put forward to make better use of user interests.
\end{itemize}
\begin{table}[ht]
\vspace{-5mm}
\setlength{\abovecaptionskip}{-0.1cm}
\setlength{\belowcaptionskip}{-0.1cm}
\caption{Performances of different MISS variants.}
    \centering
\setlength{\tabcolsep}{0.5mm}{
\small
\begin{tabular}{c|c|c|c|c|c|c}
\hline \hline
Dataset &
\multicolumn{2}{c|}{Amazon-Cds} & 
\multicolumn{2}{c|}{Amazon-Books} &
\multicolumn{2}{c}{Alipay} \\ \hline 
Model & AUC & Logloss & AUC & Logloss  & AUC & Logloss \\\hline \hline
\footnotesize{IPNN-MISS} & \textbf{0.8858} & \textbf{0.4368} & \textbf{0.9146} & \textbf{0.3778} & \textbf{0.9004} & \textbf{0.4006}   \\
\footnotesize{IPNN-MISS/F} & \underline{0.8741} & \underline{0.4506} & \underline{0.9122} & \underline{0.3880} & \underline{0.8973} & \underline{0.4044}   \\
\footnotesize{IPNN-MISS/F/U} & 0.8620 & 0.4649 & 0.8946 & 0.4116 & 0.8937 &0.4120 \\
\footnotesize{IPNN-MISS/F/L} & 0.8549& 0.4769& 0.8931 & 0.4209 & 0.8885 & 0.4214  \\
\footnotesize{IPNN-MISS/F/U/L}  & 0.8480 & 0.4829 & 0.8739 & 0.4625 & 0.8871 & 0.4220\\ 
\footnotesize{IPNN-MISS/M/F/U/L} & 0.8393 & 0.4953 & 0.8449 & 0.4933 & 0.8838 & 0.4265\\ 
\footnotesize{IPNN} & 0.8053 & 0.5364 & 0.8051 & 0.5308 & 0.8823 & 0.4299  \\

\hline
\footnotesize{DIN-MISS} & \textbf{0.8867} & \textbf{0.4357} & \textbf{0.9180} & \textbf{0.3730} & \textbf{0.9327} & \textbf{0.3295}   \\
\footnotesize{DIN-MISS/F}  & \underline{0.8813} & \underline{0.4419} & \underline{0.9136} & \underline{0.3823}  & \underline{0.9300} & \underline{0.3357}   \\
\footnotesize{DIN-MISS/F/U} & 0.8636  & 0.4642 & 0.8978 & 0.4100 &  0.9260 & 0.3446 \\
\footnotesize{DIN-MISS/F/L} & 0.8568 & 0.4807 & 0.8937 & 0.4188 & 0.9236 & 0.3499   \\
\footnotesize{DIN-MISS/F/U/L} & 0.8514 & 0.4869 & 0.8717 & 0.4642 &  0.9222 & 0.3509\\ 
\footnotesize{DIN-MISS/M/F/U/L} & 0.8429 & 0.4978 & 0.8425 & 0.4948 & 0.9188  & 0.3579  \\
\footnotesize{DIN} & 0.8055 & 0.5357  &  0.8074 & 0.5289 & 0.9098 & 0.3734  \\
\hline \hline
\end{tabular}}
\label{tab:effectiveness}
\vspace{-3mm}
\end{table}
\subsubsection{Effectiveness Analysis}
As firstly addressed in the Introduction and also reflected in the model structure, our MISS framework is built upon some important practices including the multi-interest consideration (M), the union-wise interest representation (U), the long-range interest dependencies (L), and the intra-item feature correlation (F). 
To evaluate the effectiveness of these different practices, we explore MISS with different settings. By removing some of the practices, five more MISS variants are obtained and named as MISS/F, MISS/F/U, MISS/F/L, MISS/F/U/L, and MISS/M/F/U/L respectively. All MISS variants are applied to the IPNN, DIN, and FiGNN models to verify their performances, where the resulting models are also named in an ``A''-``B'' manner.
The comparison results are presented in Table \ref{tab:effectiveness}, where the results of the FiGNN model are also omitted to save space.
As can be observed, all MISS variants bring about performance boosts to the original IPNN and DIN models, and the complete MISS framework achieves the best results. 
Therefore, we claim that all four practices (M, U, L, and F) are effective and complementary to each other, and it is necessary to adopt all of them for better performances. 
What is more, the removal of M results into the worst performance decay, revealing the importance of multi-interest modeling.

\begin{table}[ht]
\setlength{\abovecaptionskip}{-0.1cm}
\setlength{\belowcaptionskip}{-0.1cm}
\vspace{-5mm}
\caption{Performances of different Multi-Interest Extractor.}
    \centering
\setlength{\tabcolsep}{1mm}{
\small
\begin{tabular}{c|c|c|c|c|c|c}
\hline\hline
Dataset &
\multicolumn{2}{c|}{Amazon-Cds} & 
\multicolumn{2}{c|}{Amazon-Books} &
\multicolumn{2}{c}{Alipay} \\ \hline 
Extractor & AUC & Logloss &AUC & Logloss & AUC & Logloss \\\hline \hline
DIN  & 0.8055 & 0.5357  &  0.8074 & 0.5289 & 0.9098 & 0.3734  \\
MISS-SA  & 0.8042 & 0.5385 &  0.8128 & 0.5225 & 0.9092 & 0.3758  \\
MISS-LSTM & 0.8106 &  0.5299 & 0.8172 & 0.5178 & 0.9096 & 0.3753  \\
MISS-CNN & \textbf{0.8867} & \textbf{0.4357} & \textbf{0.9180} & \textbf{0.3730} & \textbf{0.9327} & \textbf{0.3295}  \\
\hline \hline
\end{tabular}}
\vspace{-2mm}
\label{tab:Effect_of_interest_extraction}
\end{table}
\subsubsection{Multi-Interest Extractor Analysis}
\begin{figure}[ht]
\centering
\setlength{\abovecaptionskip}{-0.2cm}
\vspace{-3mm}
\noindent\makebox[0.5\textwidth][c]{\includegraphics[scale=0.25]{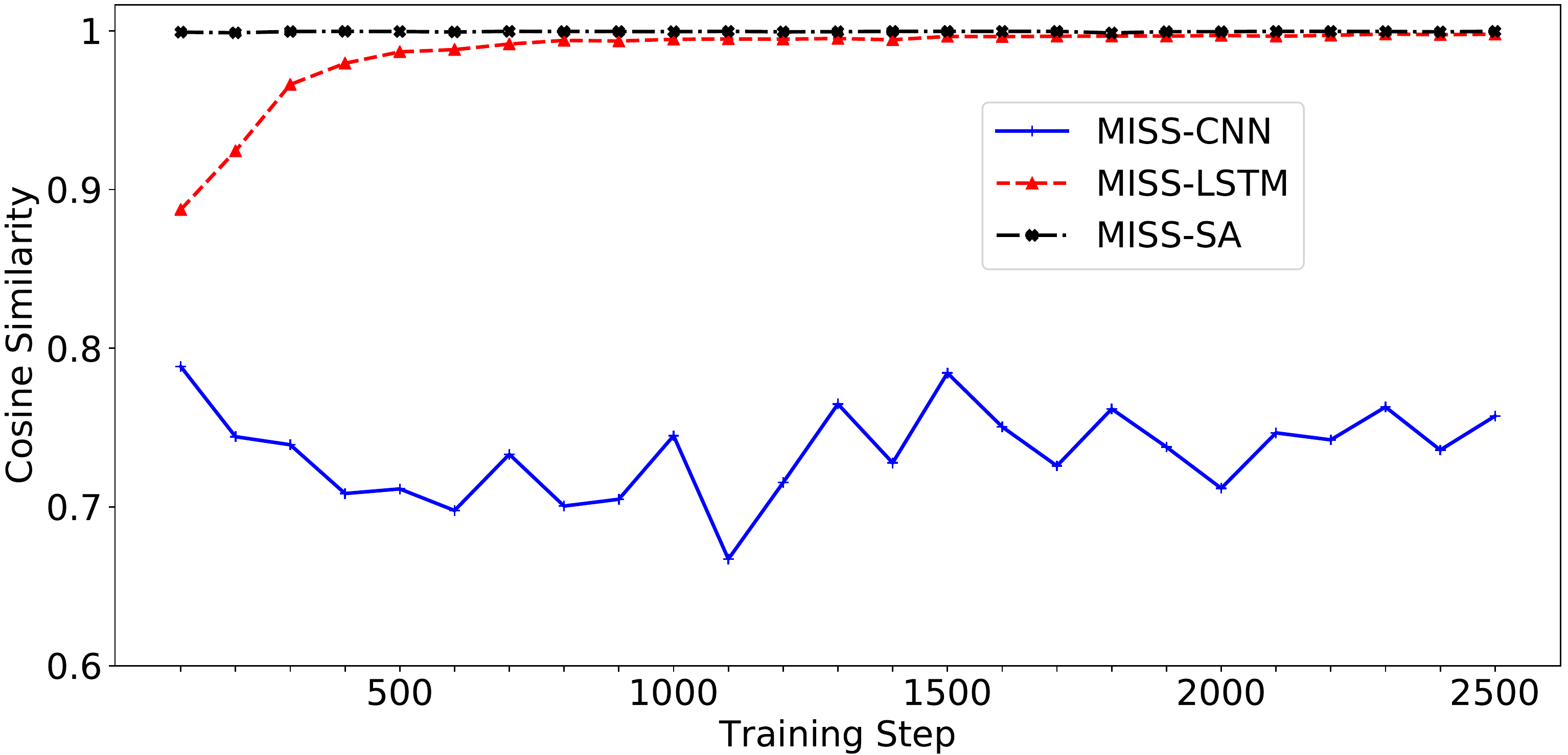}}
\caption{Similarity analysis in Amazon-Cds.}
\label{fig:similarity-comparison}
\vspace{-5mm}
\end{figure}

To verify the rational of our $\text{MIE}(\cdot)$ design formulated in Equation (\ref{eq3}-\ref{equ:mie_detail}), we compare the performances of our proposed CNN module with self-attention \cite{vaswani2017attention} and LSTM \cite{hochreiter1997long} for multi-interest extraction, and the resulting models are named as MISS-CNN (the same model as MISS), MISS-SA, MISS-LSTM respectively.
Table VIII summarizes the experimental results.
We can see that the CNN extractor achieves the best performances on all datasets.
For an in-depth analysis of these extractors, we further visualize the cosine similarity scores between the generated pairs of views from these interest representations on Figure \ref{fig:similarity-comparison}.
Each training step on the x-axis in Figure \ref{fig:similarity-comparison} corresponds to a batch of training samples fed at that step, and the average similarity score among training batches are reported.
As we can see, the similarity scores of MISS-SA and MISS-LSTM are close to 1, thus the generated pairs hardly provide any useful information for contrastive learning.
The reason may be as follows. LSTM learns the characteristics of the whole historical behavior sequence, and the histories of two adjacent items of the sequence only differ by one item, so the representations learned for the two adjacent items via LSTM are highly similar. Self-attention based method aggregates all behaviors to generate interest representations, and hence learns similar representations for adjacent items. In comparison, CNN based model considers a sliding window of the past history, and the size of the sliding window is small (at most 3 or 4 in our experiments), so differing by one most recent item will make a notable difference in the representation. This is evidenced by the similarity scores of our proposed CNN model, which are in the range of 0.7 and 0.8. The representations of interest at adjacent timestamps are similar but also distinguishable for contrastive learning. This validates the superiority of using CNN compared to LSTM or self-attention in our problem.

\begin{figure*}[ht]
\centering
\setlength{\abovecaptionskip}{-0.2cm}
{\includegraphics[scale=0.5]{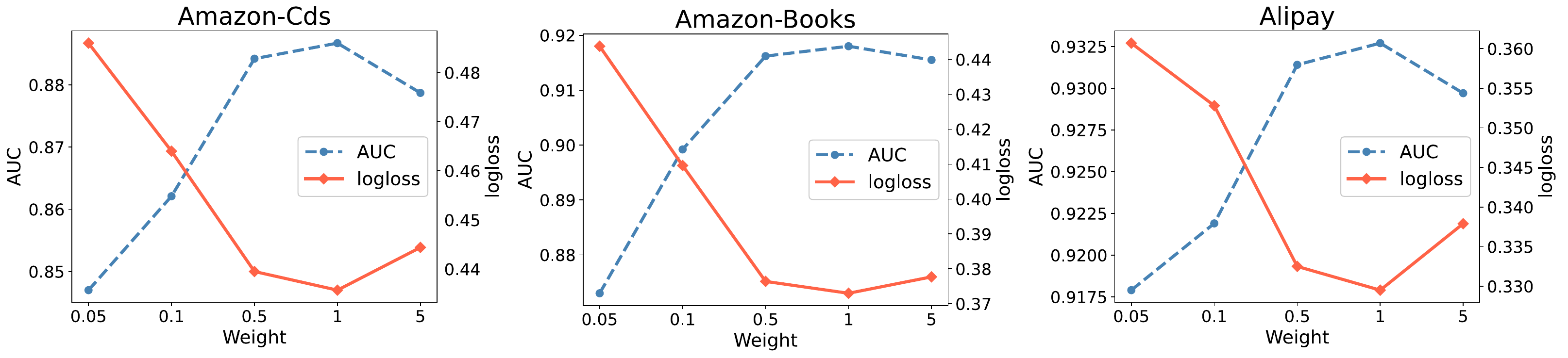}}
\caption{Performances of MISS w.r.t. different weights assigned to the SSL losses.} 
\label{fig:weight}
\vspace{-2mm}
\end{figure*}

\begin{figure*}[ht]
\centering
\setlength{\abovecaptionskip}{-0.2cm}
{\includegraphics[scale=0.5]{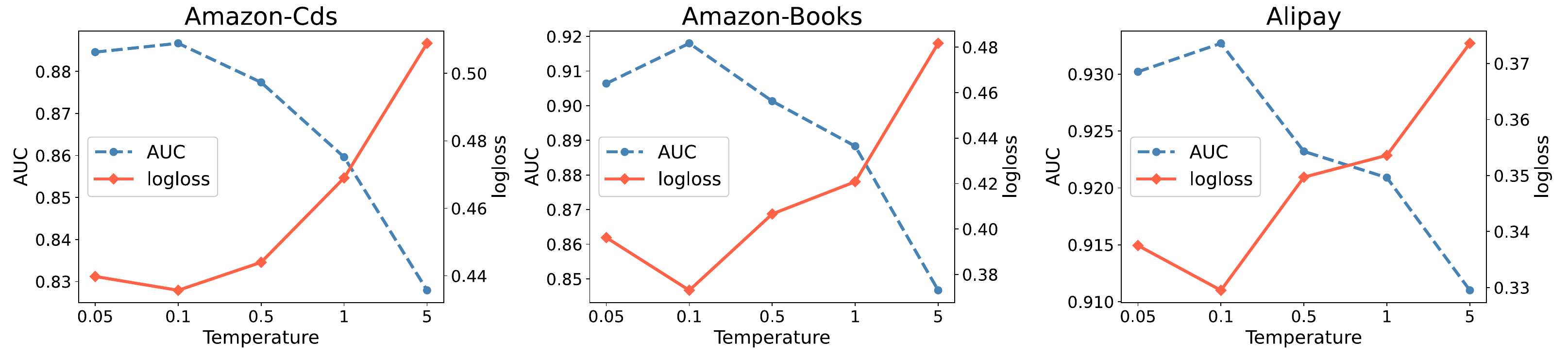}}
\caption{Performances of MISS w.r.t. different temperature parameter values.} 
\label{fig:temperature}
\vspace{-5mm}
\end{figure*}
\subsection{Model Training Analysis}
During training, our proposed MISS framework has several key hyper-parameters that may affect the performances, and so do the multi-task training strategies. In this section, we first investigate the importance and sensitivity of these hyper-parameters by changing one hyper-parameter while fixing the others. 
After that, different training strategies of the two losses are also compared.

\subsubsection{Impact of the loss weight}
The final loss function of MISS in Equation (\ref{equ:final loss}) is a combination of the CTR prediction loss and the SSL losses. 
Figure \ref{fig:weight} shows the CTR prediction performances under different loss weights where larger weights indicate stronger contributions of the SSL losses.
We can observe that the performances grow stably with the increase of the loss weight at the beginning.
However, when the weight grows bigger than 1, performance degradation happens.
Thus the SSL losses should not dominate the training process. 
In other words, the SSL part takes the auxiliary role for CTR prediction, and the model can be biased when it is overemphasized. 

\subsubsection{Impact of the softmax temperature}
The softmax temperature parameters in Equation (\ref{equ:self-loss-1}) and Equation (\ref{equ:self-loss-2}) tune the distribution of the SSL losses. 
A large temperature value will draw close the predictions of positive and negative samples in SSL losses, thus weakens the supervision signals in training.
We analyze how different temperature parameter values affect the model performances, and the results are illustrated in Figure \ref{fig:temperature}. 
With the growth of the temperature value, performances on all three datasets increase first and then decrease. 
The turning point is 0.1 for both metrics on all datasets. 
With such a small temperature value (significantly less than 1), the supervision signals get strengthened during training as the positive and negative SSL samples are better discriminated. 
In other words, discriminating positive and negative samples benefits the model performances, which accords with our motivations.

\begin{table}[ht]
\setlength{\abovecaptionskip}{-0.1cm}
\setlength{\belowcaptionskip}{-0.1cm}
\vspace{-5mm}
\caption{Performances of different MISS training strategies.}
    \centering
\setlength{\tabcolsep}{1mm}{
\small
\begin{tabular}{c|c|c|c|c|c|c}
\hline\hline
Dataset &
\multicolumn{2}{c|}{Amazon-Cds} & 
\multicolumn{2}{c|}{Amazon-Books} &
\multicolumn{2}{c}{Alipay} \\ \hline 
Model & AUC & Logloss & AUC & Logloss  & AUC & Logloss \\\hline \hline
DIN & 0.8055 & 0.5357  &  0.8074 & 0.5289 & 0.9098 & 0.3734  \\
MISS-Joint & \textbf{0.8867} & \textbf{0.4357} & \textbf{0.9180} & \textbf{0.3730} & \textbf{0.9327} & \textbf{0.3295}  \\
MISS-Pre & 0.8848 & 0.4381 & 0.9170 & 0.3746 & 0.9313 & 0.3328  \\
\hline \hline
\end{tabular}}
\label{tab:Effect of learning strategy}
\vspace{-5mm}
\end{table}

\subsubsection{Training strategies} 
There are two learning targets in our MISS framework, i.e., CTR prediction and self-supervised learning. 
During training, different multi-task learning strategies can be adopted to optimize the two targets. 
Right here, we compare and analyze the most widely used joint learning and pre-training strategies. 
Table \ref{tab:Effect of learning strategy} gives the analysis results, where DIN is used as the backbone model. 
The MISS model trained with joint learning is denoted as MISS-Join, while MISS-Pre learns the CTR prediction target based on the pre-trained embeddings by MISS.
Both MISS-Join and MISS-Pre achieve better performances than DIN, and MISS-Joint perform even better than MISS-Pre. 
In the joint end-to-end training, complementary supervision signals are shared across the two targets, resulting into mutual enhancements that are beyond the reach of pre-training.

\subsection{Case Study}
In this part, we verify that our model can effectively alleviate the label sparsity and label noise problems.

\subsubsection{Label Sparsity Analysis}
As explained in the Introduction, CTR models easily suffer from the label sparsity issue.
To verify our model's effectiveness in alleviating label sparsity, we down-sample the original training set with sampling rate (SR) 90\% and 80\%, while the validation and testing sets stay unchanged.
Notice that the 100\% sampling rate means using the original training set.
Table \ref{tab:label_sparsity} shows the performances with different SR, where the results on the Alipay dataset are omitted for space limitation. 
We omit the results on the Alipay dataset for space limitation, which have similar trends.
It can be found that the performance drops when the labels become sparse (SR decreases), while the relative improvement (RI) gets larger.
Thus our MISS model can effectively alleviate the label sparsity problem.
\begin{table}[t]
\setlength{\abovecaptionskip}{-0.1cm}
\setlength{\belowcaptionskip}{-0.1cm}
    \centering
\setlength{\tabcolsep}{0.5mm}
\small
\caption{AUC Scores of CTR prediction with different sampling rate.}
{
\begin{tabular}{c|c|c|c|c|c|c}
\hline \hline
Dataset &
\multicolumn{3}{c|}{Amazon-Cds} & 
\multicolumn{3}{c}{Amazon-Books} \\ \hline 
SR & DIN & DIN-MISS & RI & Din & DIN-MISS & RI \\\hline \hline
80\% &  0.7913 & 0.8779 & \textbf{10.94\%} & 0.7932 & 0.9107 & \textbf{14.81\%}  \\
90\% &  \underline{0.7988} & \underline{0.8814} & \underline{10.34\%} & \underline{0.7998} & \underline{0.9152} & \underline{14.43\%} \\
100\%&  \textbf{0.8055} & \textbf{0.8867} & 10.08\% & \textbf{0.8074} & \textbf{0.9180} & 13.70\% \\
\hline \hline
\end{tabular}}
\label{tab:label_sparsity}
\vspace{-5mm}
\end{table}

\begin{table}[t]
\setlength{\abovecaptionskip}{-0.1cm}
\setlength{\belowcaptionskip}{-0.1cm}
    \centering
\setlength{\tabcolsep}{0.5mm}
\small
\caption{AUC Scores of CTR prediction with different label noise rate.}
{
\begin{tabular}{c|c|c|c|c|c|c}
\hline \hline
Dataset &
\multicolumn{3}{c|}{Amazon-Cds} & 
\multicolumn{3}{c}{Amazon-Books} \\ \hline 
NR &  DIN & DIN-MISS & RI & DIN & DIN-MISS & RI \\\hline \hline
0\% &  \textbf{0.8055} & \textbf{0.8867} & 10.08\% & \textbf{0.8074} & \textbf{0.9180} & 13.70\%  \\
10\% &  \underline{0.7768} & \underline{0.8652} & \underline{11.38\%} & \underline{0.7775} & \underline{0.8877} & \underline{14.16\%} \\
20\%&  0.7413 & 0.8331 & \textbf{12.38\%} & 0.7384 & 0.8678 & \textbf{17.52}\%\\
\hline \hline
\end{tabular}}
\label{tab:label_noise}
\vspace{-5mm}
\end{table}
\subsubsection{Label Noise Analysis}
Besides label sparsity, the label noise problem can also be well solved by our proposed MISS model.
To check the robustness of MISS to label noise, noises are imposed on the training set by randomly swapping the labels at an indicated proportion (10\% and 20\%) of samples, while the validation and testing sets stay unchanged. 
Notice that 0\% noise rate (NR) means using the original training set.
Due to space limitation, only the results on Amazon-Cds and Amazon-Books datasets are demonstrated in Table \ref{tab:label_noise}.
It is obvious that the relative improvement of DIN-MISS over DIN grows more significant when NR increases.
In other words, MISS shows good robustness to label noise.

\section{Conclusion}
In this paper, we proposed a Multi-Interest Self-Supervised learning (MISS) framework for the CTR prediction task.
In view of the multi-interest characteristics of user behaviors, a CNN-based multi-interest extractor component was proposed to learn the hidden interests while considering both point-wise and union-wise interest representations. 
Further, another CNN-based multi-feature extractor was also proposed to utilize both inter-item and intra-item interest correlations at the fine-grained feature level.
With the help of two random selection functions, augmented views of interest representations can be extracted in consideration of both short-range and long-range interest dependencies.
Based on the augmented views of interest representations, two contrastive learning losses effectively transforms interest correlation knowledge into self-supervision signals. 
In this way, not only the label sparsity issue gets alleviated by the self-supervision signals, but also the model robustness gets enhanced to shield label noise. 
Extensive experimental results on three large-scale datasets verify the effectiveness of the proposed MISS framework.


\bibliographystyle{IEEEtran}
\bibliography{main}

\end{document}